\documentclass[11pt,preprint,superscriptaddress,aps,showkeys,nofootinbib]{revtex4}

\usepackage{amssymb,amsmath,amsfonts}
\usepackage{graphicx}
\usepackage{graphics}
\usepackage{subfig}
\usepackage{eepic,epsfig}
\usepackage{indentfirst}
\usepackage[utf8]{inputenc}
\usepackage[T1]{fontenc}
\usepackage{hyperref}
1.60cm \makeatletter \@addtoreset{equation}{section}

\makeatother

\begin{document}
\title{Casimir energy and topological mass for a massive scalar field with Lorentz violation}
\author{M. B. Cruz}
\email{messiasdebritocruz@gmail.com}
\affiliation{Departamento de F\'{\i}sica, Universidade Federal de Campina Grande\\
 Caixa Postal 882, 58428-830, Campina Grande, Para\'{\i}ba, Brazil}
\author{E. R. Bezerra de Mello}
\email{emello@fisica.ufpb.br}
\affiliation{Departamento de F\'{\i}sica, Universidade Federal da Para\'{\i}ba\\
 Caixa Postal 5008, 58051-970, Jo\~ao Pessoa, Para\'{\i}ba, Brazil}
\author{H. F. Santana Mota}
\email{hmota@fisica.ufpb.br}
\affiliation{Departamento de F\'{\i}sica, Universidade Federal da Para\'{\i}ba\\
 Caixa Postal 5008, 58051-970, Jo\~ao Pessoa, Para\'{\i}ba, Brazil}

\begin{abstract}
A Lorentz symmetry violation aether-type theoretical model is considered to investigate the Casimir effect and the generation of topological mass associated with a self-interacting massive scalar fields obeying Dirichlet, Newman and mixed boundary conditions on two large and parallel plates. By adopting the path integral approach we found the effective potential at one- and two-loop corrections which provides both the energy density and topological mass when taken in the ground state of the scalar field. We then analyse how these quantities are affected by the Lorentz symmetry violation and compare the results with previous ones found in literature.   
\end{abstract}
\keywords{Casimir effect, Lorentz violation, scalar fields, loop corrections, topological mass.}

\maketitle

\section{Introduction} 
\label{introd}

The Casimir effect was predicted by H. B. Casimir in 1948 \cite{casimir:1948} and, although not with a great precision, experimentally verified teen years later by M. J. Sparnnaay \cite{Sparnaay:1958}. Since then it has been confirmed by several high precision experiments \cite{Bressi:2002fr, Lamoreaux:1996wh, PhysRevLett.81.5475, mohideen1998precision, MOSTEPANENKO2000, PhysRevA.78.020101, PhysRevA.81.052115}, leading currently to one of the most interesting topic of research. The Casimir effect consists in a direct manifestations of the existence of quantum fluctuations of the vacuum and was noted to arise for the first time when considering two parallel conducting plates placed very close to each other in the vacuum, separated by a very small distance when compared with the plates dimensions. In this case, the theoretical prediction and experimental observation that the two plates attract each other \cite{casimir:1948} was not credited to the gravitational or electromagnetic forces, but to the modifications of the quantum vacuum fluctuations of the electromagnetic field by the presence of the plate. The gravitational interaction between the plates is far too weak to be observed while the electromagnetic interaction is absent since the plates are neutral. 

Other quantum-relativistic fields, such as scalar and fermion fields, can also present modifications in the quantum fluctuations of their vacua by some sort of boundary condition, leading to a Casimir-like effect. The formal and standard approach to investigate the Casimir interactions is in the realm of quantum field theory, which is based on the assumption that the Lorentz symmetry is preserved. However, one may well assume other scenarios where the Lorentz symmetry is violated which is normally the case in models that look for probing high-energy phenomena. This, in fact, has been done from both theoretical and experimental point of views. In the context of several of these Lorentz symmetry violation scenarios the spacetime becomes anisotropic in one (or more) direction, including time, and inevitably the quantum field whose modes propagates in it has its energy spectrum modified. Lorentz symmetry violation in string theory can be found in Ref. \cite{kostelecky:1989} and in low-energy scale scenarios in Refs. \cite{alexey:2002,carlson:2001,hewett:2001,bertolami:2003, kostelecky:2003,anchordoqui:2003,bertolami:1997, alfaro:2000,alfaro:2002}. In these contexts, Casimir energy has been considered in Refs. \cite{Obousy:2008xi, Obousy:2008ca} and \cite{ Martin-Ruiz:2016ijc, Martin-Ruiz:2016lyy, Escobar:2020pes}, respectively. Therefore, with the great number of theoretical works, it is natural that the search for Lorentz symmetry violations also acquire experimental interest and, in this sense, the Casimir effect becomes an even more interesting topic to study since it can be related with Lorentz symmetry violation models. 

Although the Casimir effect is more often calculated in terms of the zero-point energy of a quantized field, this effect can also be investigated by adopting the path integral formalism for quantum field theory, developed by Jackiw \cite{jackiw:1686}, in which an effective potential, presented in terms of loop-expansions, allows us to obtain the energy density as well as generation of topological mass.\footnote{In fact both calculations are divergent. The standard procedure to renormalize them and obtain finite results is through the use of Riemann zeta-function.}  Studies of radiative corrections for the Casimir energy were reported in Refs. \cite{wieczorek:1986,barone:2004,robaschik:1987} and in \cite{toms:1980:928,toms:1980:2805,toms:1980:334357}. In the latter, generation of topological mass for a self-interacting massless scalar field obeying different boundary conditions on two large and parallel plates was also considered. By following the same line of investigation as in Ref. \cite{toms:1980:2805}, in the present work, we study the loop expansions to the Casimir energy and generation of topological mass for a self-interacting massive and massless scalar fields subject to Dirichlet, Newman and mixed boundary condition in the context of a CPT even aether-type Lorentz symmetry violation model \cite{carroll:2008,chatrabhuti:2009,gomes:2010}. 

This paper is organized as follows: In section \ref{casim_effect} we briefly describe the theoretical model  that we want to investigate, which consists  of a self-interacting massive scalar field in a CPT even aether-type Lorentz symmetry violation approach. We then calculate the one- and two-loop radiative corrections to the Casimir energy and generation of topological mass admitting that the scalar field obeys Dirichlet, Newman and mixed boundary conditions on two large and parallel plates. Because these calculations are divergent, we adopt the Riemann zeta-function renormalization procedure to provide finite and well defined results. Finally, in section \ref{concl} we present our conclusions. Throughout the paper we use natural units $\hbar = c = 1$ and metric signature $(-,+,+,+)$.


\section{Loops corrections and generation of topological mass}
\label{casim_effect}
\subsection{Theoretical Model}

We first introduce the aether-type Lorentz symmetry violation model  that we want to consider to investigate the vacuum energy and generation of topological mass. The model is composed by a self-interacting scalar field that presents a CPT even and aether-like Lorentz violation term implemented by direct coupling between the derivative of the field with an external constant $4-$vector. (For a more detailed review see \cite{carroll:2008,gomes:2010}). The model is described by the action below,
\begin{equation}
 \label{model_action}
 \mathcal{S}(\phi) = \int_{\mathcal{M}} d^4x \mathcal{L}(x) \ ,
\end{equation}
where $\mathcal{M}$ is a flat manifold and $\mathcal{L}$ is a Lagrangian density, given by
\begin{equation}
 \label{lagrang_density}
 \mathcal{L}(x) = - \frac{1}{2} \big(\partial_{\mu}\phi \big)\big(\partial^{\mu}\phi \big) + \frac{1}{2} \chi \big(u \cdot \partial \phi \big)^2 - U(\phi) \ .
\end{equation}
In the above expression, the scalar field of mass $m$ is represented by $\phi(x)$. The $4$-vector, $u^{\mu}$, is responsible for a privileged direction in spacetime and the dimensionless parameter $\chi$, which codifies the Lorentz violation, is much smaller than unity. The last term on the r.h.s of \eqref{lagrang_density} is the classical potential $U(\phi)$, which for a massive and self-coupling, $\lambda\phi^4$ theory, is given by
\begin{equation}
 \label{potencial_u}
 U(\phi) = \frac{m^2\phi^2}{2} + \frac{\lambda \phi^4}{4!} + \frac{\phi^4}{4!}\delta_1 + \frac{\phi^2}{2}\delta_2 + \delta_3 \ ,
\end{equation}
where the parameters $\delta_1$, $\delta_2$ and $\delta_3$ correspond to the renormalization constants of the theory and will be determined later. Before we proceed, we want to make clear that the analysis we want to develop in this paper will take into consideration the $4$-vector, $u^{\mu}$, in two types: timelike and spacelike. The timelike component is represented by $u^t=(1,0,0,0)$ while the spacelike is represented by $u^x=(0,1,0,0)$ if the privileged direction is in the $x$-axis, $u^y=(0,0,1,0)$ if the privileged direction is in the $y$-axis and $u^z=(0,0,0,1)$ if the privileged direction is in the $z$-axis.

In order to adopt the path integral approach described in detail in Ref. \cite{toms:1980:2805} we need to allow the field $\phi(x)$ to fluctuate about a fixed background field, $\Phi$, with its quantum fluctuations represented by $\varphi$. Thus, after performing a Wick rotation $(t\rightarrow -it)$ in the Lorentzian action \eqref{model_action} and define an Euclidean one we can make use of the generating function of the one-particle-irreducible Green function \cite{toms:1980:2805}. This provides a description in terms of a $\Phi$-dependent effective potential which, up to two-loop corrections, is written as 
\begin{eqnarray} 
V_{\text{eff}}(\Phi) = V_{\text{cl}}(\Phi) + V^{(1)}(\Phi) + V^{(2)}(\Phi),
\label{effP}
\end{eqnarray}
where $V_{\text{cl}}(\Phi)= U(\Phi)$ is the tree-level (classical) contribution to the effective potential in a flat manifold, $V^{(1)}(\Phi)$ and $V^{(2)}(\Phi)$ are the one- and two-loop correction contributions, respectively. Note that we have performed a linear expansion of $\phi$ ($\phi\rightarrow\Phi + \varphi$) about the classical field, $\Phi$. The two-loop contribution in the last term on the r.h.s of \eqref{effP} is a contribution of two graphs to the effective Euclidian action \cite{toms:1980:2805}. We will postpone to the next sections how to calculate it, for each case we consider. 

As to the one-loop contribution to the effective potential, we will follow the same method of \cite{toms:1980:2805}, which is to define this contribution in terms of the Riemann zeta function $\zeta(s)$, i.e.,
\begin{equation} \label{one_loop_potential_DN_t}
 V^{(1)}(\Phi) = - \frac{1}{2 \text{vol} (\text{E})} \bigg{[}\zeta^{\prime}(0) + \zeta(0) \text{ln} \left(\mu^2\right) \bigg{]} \ ,
\end{equation}
where $\text{vol(E)}$ is the Euclidian volume, $\zeta^{\prime}(s)$  the derivative of the zeta function with respect to the parameter $s$ and the term $\zeta(0)\text{ln}(\mu^2)$ is to be removed by renormalization.\footnote{The parameter $\mu$ is associated with a measure on the space of function \cite{toms:1980:2805}.} As it is known, the (generalized) Riemann zeta function, $\zeta(s)$, is defined as 
\begin{equation}  \label{zet_funct_DN_t_0}
 \zeta(s) = \sum_{\beta} \Lambda_{\beta}^{-s} \ ,
\end{equation}
where $\Lambda_{\beta}$ is the spectrum of eigenvalues associated with a self-adjoint elliptic operator, which in our case is given by

\begin{eqnarray}
\Delta = -\partial_{\mu}\partial^{\mu} + \chi u^{\mu}u^{\nu}\partial_{\mu}\partial_{\nu} + m^2 + \frac{\lambda\Phi^2}{2}.
\label{operator}
\end{eqnarray}
Note that $\beta$ stands for the set of quantum numbers associated with the quantum field eigenfunction, $\varphi$, of the operator, $\Delta$. Although the zeta function \eqref{zet_funct_DN_t_0} is defined in terms of the complex parameter $s$, for $\text{Re}(s) > 1$, an analytic continuation to the whole complex plane can be obtained for it, including in $s = 0$. 

The renormalization condition that enable us to eliminate the term $\zeta(0)\text{ln}(\mu^2)$ in \eqref{one_loop_potential_DN_t} is considered in analogy to Coleman-Weinberg and should fix the coupling-constant \cite{coleman,toms:1980:2805}. This condition is written as 
\begin{eqnarray}
\frac{d^4V_{\text{eff}}}{d\Phi^4}\bigg|_{\Phi=0}=\lambda.
\label{ren1}
\end{eqnarray}
As we will see, this condition will fix the renormalization constant $\delta_1$.
\begin{figure}[h!]
 \includegraphics[scale=0.7]{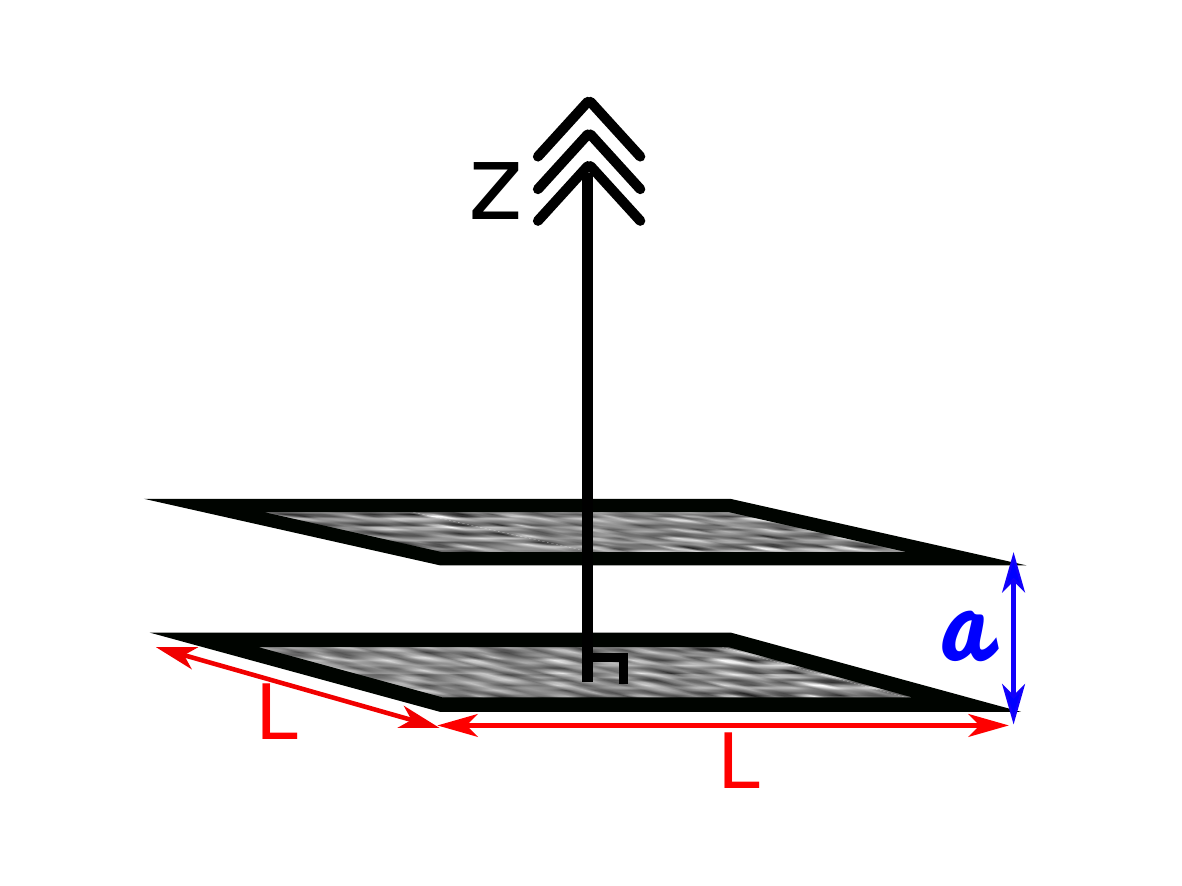}
 \label{par_plat}
 \caption{Schematic configuration for the two parallel plates with area $L^2$  separated by a distance $a$ ($a \ll L$).}
 \label{figure1}
\end{figure}

On the other hand, the condition that makes possible to obtain a topological mass is given by
\begin{equation}
\frac{d^{2}V_{\text{eff}}}{d\Phi^2}\bigg|_{\Phi=0}=m^2.
\label{ren2}
\end{equation} 
This condition fix the renormalization constant $\delta_2$ and also provides the topological mass when using the renormalized effective potential, as we will see. Note that $\Phi=0$ is the value that minimizes the effective potential and represents the minimum of the potential only if it obeys the extremum condition
\begin{equation}
\frac{dV_{\text{eff}}}{d\Phi}\bigg|_{\Phi=0}=0,
\end{equation}
leading to Eq.(\ref{ren2}) being positive. 

Moreover, in order to find the constant $\delta_3$ we also need to use an additional renormalization condition which is given by 
\begin{eqnarray}
 V_{\text{eff}}\big|_{\Phi=0}=0.
 \label{ren3}
 \end{eqnarray}

From now on in our discussion we will assume that the quantum field $\varphi$ is confined between two large parallel plates, as it is shown in Fig.\ref{figure1}. The quantum field $\varphi$ is an eigenfunction of the self-adjoint elliptic operator \eqref{operator}, with eigenvalues $\Lambda_{\beta}$. In this sense, the eigenvalues we are interested in are the ones obtained by requiring that $\varphi$ must satisfy specific boundary conditions on the plates placed at $z=0$ and $z=a$ for all cases of $4$-vector $u^{\mu}$: timelike and spacelike.

\subsection{Dirichlet and Neumann boundary conditions} \label{DN_bound_cond}

We will start by considering the case where the scalar field, $\varphi$, satisfies Dirichlet and Neumann boundary conditions on the plates, respectively, that is,
 \begin{equation}
 \label{dir_bound}
  \varphi(x)\big{|}_{z=0} = \varphi(x)\big{|}_{z=a}, \qquad\text{and}\qquad \frac{\partial \varphi(x)}{\partial z}\bigg{|}_{z=0} = \frac{\partial \varphi(x)}{\partial z}\bigg{|}_{z=a} \ .
 \end{equation}
The complete set of normalized solutions of the scalar field, $\varphi$, under these conditions have been reported, for instance, in Ref. \cite{cruz:2017}.  These solutions provide the following eigenvalues of the operator \eqref{operator}:
\begin{eqnarray}
\Lambda_{\beta} = k_t^2 + k^2 + \frac{n^2\pi^2}{a^2}  -\chi u^{\mu}u^{\nu}k_{\mu}k_{\nu} + m^2 + \frac{\lambda\Phi^2}{2},
\label{ev1}
\end{eqnarray}
where $k_{\mu}=(k_t ,k_x, k_y, k_z)$ are the four-momentum components, $k^2=k_x^2+k_y^2$ and $k_z=n\pi/a$, for $n=1,2,3,...$\ . Hence, the set of quantum numbers is $\beta = (k_t,k_x,k_y,n)$. Note that $k_t,k_x$ and $k_y$ are continuum quantum numbers. In addition we want to point out that for constant time-like vector, $u^t$, in \eqref{ev1}, we have to consider the Euclidean version of the zero-component of the four-momentum, i.e., we have to take $k_t\to- ik_t$, in the term associated with the Lorentz violating parameter, $\chi$. 
%
\subsubsection{Timelike vector}
%
We want to begin considering the timelike type of the 4-vector $u^{\mu}$, in which case, $u^{t}=(1,0,0,0)$, meaning that the privileged direction chosen to have the Lorentz symmetry violated is the time one. This leads to the eigenvalues \eqref{ev1} to be written as
\begin{equation} 
 \label{eigenv_timelike_DN}
 \Lambda_{\beta} = (1+\chi)k_t^2 + k^2 + \frac{\pi^2 n^2}{a^2} + m^2 + \frac{\lambda \Phi^2}{2} \ .
\end{equation}
The set of eigenvalues in Eq. \eqref{eigenv_timelike_DN} allow us to build the zeta function by using the definition \eqref{zet_funct_DN_t_0}. This provides 
\begin{equation} \label{zeta_DN_t_0}
 \zeta(s) = \frac{V_3}{(2\pi)^3} \sum_{n=1}^{\infty} \int d^3k \left[(1+\chi)k_t^2 + k^2 + \frac{\pi^2 n^2}{a^2} + m^2 + \frac{\lambda}{2}\Phi^2\right]^{-s} \ ,
\end{equation}
where $V_3$ is a continuum volume associated with the coordinates $t,x,y$ and $d^3k=dk_tdk_xdk_y$. After defining a new variable $\kappa_t=\sqrt{1+\chi}k_t$, the integrals in $\kappa_t \ , \ k_x$ and $k_y$ can be performed by using the identity
\begin{equation} 
 \label{identity}
\frac{1}{\omega^{2s}} = \frac{1}{\Gamma(s)}\int_0^{\infty}d\tau\tau^{2s-1}e^{-\omega^2\tau^2}.
\end{equation}
Consequently, \eqref{zeta_DN_t_0} becomes 
\begin{equation}
 \label{zet_funct_DN_t_1}
 \begin{aligned}
 \zeta(s) =  \frac{V_3}{(2\pi)^3}\frac{\pi^{3/2}}{\sqrt{1+\chi}}\frac{\Gamma(s-3/2)}{\Gamma(s)}w^{3-2s}\sum_{n=1}^{\infty}\left(n^2+\nu^2\right)^{3/2-s} \ ,
 \end{aligned}
\end{equation}
where
\begin{equation}
 \begin{aligned}
  \nu^2 \equiv \frac{\lambda \Phi^2}{2w^2} + \frac{m^2}{w^2} \ \ \ \ \ \ \ \ \text{and} \ \ \ \ \ \ \ \ w \equiv \frac{\pi}{a} \ .
 \end{aligned}
\end{equation}
The sum in $n$ present in Eq. \eqref{zet_funct_DN_t_1} can be worked out by making use of the Epstein-Hurwitz zeta function  \cite{grad}:
\begin{equation}
 \label{EH_funct}
 \begin{aligned}
  \zeta_{EH}(s,\nu) & \equiv \sum_{n=1}^{\infty} \left(n^2+\nu^2\right)^{-s} \\ &= -\frac{\nu^{-2s}}{2} + \frac{\pi^{1/2}}{2}\frac{\Gamma(s-1/2)}{\Gamma(s)}\nu^{1-2s} +\frac{2^{1-s}(2\pi)^{2s-1/2}}{\Gamma(s)} \sum_{n=1}^{\infty} n^{2s-1} f_{(s-1/2)}(2\pi n \nu), 
 \end{aligned}
\end{equation}
where the function $f_{\gamma}(x)$ is defined in terms of the modified Bessel functions \cite{grad}, $K_{\gamma}(x)$, by the following relation:
\begin{equation}
 f_{\gamma}(x) \equiv \frac{K_{\gamma}(x)}{x^{\gamma}}.
\end{equation}
Thus, using \eqref{EH_funct} in Eq. \eqref{zet_funct_DN_t_1} we get
\begin{equation}
 \label{zet_funct_DN_t_2}
 \begin{aligned}
  \zeta(s) = \frac{V_3}{(2\pi)^3} \frac{1}{\sqrt{1+\chi}} \bigg{[} & -\frac{\pi^{3/2} w^{-2\bar{s}} \nu^{-2\bar{s}}}{2} \frac{\Gamma(\bar{s})}{\Gamma(\bar{s}+3/2)} + \frac{\pi^{1/2} \pi^{3/2} w^{-2\bar{s}}}{2}\frac{\Gamma(\bar{s}-1/2)}{\Gamma(\bar{s}+3/2)}\nu^{1-2\bar{s}} \\ & + \frac{2^{1-\bar{s}}(2\pi)^{2\bar{s}-1/2} \pi^{3/2} w^{-2\bar{s}}}{\Gamma(\bar{s}+3/2)} \sum_{n=1}^{\infty} n^{2\bar{s}-1} f_{(\bar{s}-1/2)}(2\pi n \nu) \bigg{]} \ ,
 \end{aligned}
\end{equation}
where $\bar{s}=s-3/2$. We should note that in order to obtain the one-loop correction to the effective potential we have to take the limit, $s\rightarrow 0$, which means  $\bar{s}\rightarrow -3/2$. Consequently, Eq. \eqref{zet_funct_DN_t_2} provides
\begin{equation} \label{zet_0_D_t}
 \zeta(0) = \frac{V_3 a}{2(2\pi)^4} \frac{\pi^2 b^4}{\sqrt{1+\chi}},
\end{equation}
and
\begin{equation} \label{zet_prim_D_t}
 \begin{aligned}
  \zeta^{\prime}(0) = \frac{V_3 a}{(2\pi)^3} \left[-\frac{2\pi^2}{3}\frac{b^3}{\sqrt{1+\chi}a} + \frac{3}{8}\frac{\pi b^4}{\sqrt{1+\chi}} - \frac{\pi}{2} \frac{b^4 \text{ln}(b)}{\sqrt{1+\chi}} + \frac{2\pi b^2}{\sqrt{1+\chi}a^2}\sum_{n=1}^{\infty} \frac{K_2\left(2ban\right)}{n^2} \right],
 \end{aligned}
\end{equation}
where the parameter $b$ is defined as
\begin{equation}
 b \equiv \sqrt{\frac{\lambda \Phi^2}{2} + m^2} \ .
\end{equation}
Hence, substituting the results in Eqs. \eqref{zet_0_D_t} and \eqref{zet_prim_D_t} into \eqref{one_loop_potential_DN_t}, we find the one-loop correction to effective potential, that is,
\begin{equation} \label{1_loop_potent_DN_t}
 \begin{aligned}
  V^{(1)}(\Phi) &= \frac{b^3}{24 \pi \sqrt{1+\chi} a} - \frac{3 b^4}{128 \pi^2 \sqrt{1+\chi}} + \frac{b^4 \text{ln}\left(\frac{b^2}{\mu^2}\right)} {64 \pi^2 \sqrt{1+\chi}} \\ & \ \ \ - \frac{b^2}{8 \pi^2 \sqrt{1+\chi} a^2} \sum_{n=1}^{\infty}\frac{K_2\left(2ban\right)}{n^2} .
 \end{aligned}
\end{equation}
This allow us to write the effective potential \eqref{effP} up to one-loop correction as
\begin{equation} \label{effec_potent_DN_t}
 \begin{aligned}
  V_{\text{eff}}(\Phi) &= \frac{m^2\Phi^2}{2} + \frac{\lambda \Phi^4}{4!} + \frac{ \Phi^2}{2}\delta_2 + \frac{\Phi^4}{4!}\delta_1 + \delta_3 + \frac{b^3}{24 \pi \sqrt{1+\chi} a} - \frac{3 b^4}{128 \pi^2 \sqrt{1+\chi}} \\ & \ \ + \frac{b^4 \text{ln}\left(\frac{b^2}{\mu^2}\right)}{ 64\pi^2 \sqrt{1+\chi}} - \frac{b^2}{8 \pi^2 \sqrt{1+\chi} a^2} \sum_{n=1}^{\infty}\frac{K_2\left(2ban\right)}{n^2} .
 \end{aligned}
\end{equation}

The effective potential above still needs to be renormalized, requiring that we find the renormalization constants $\delta_1$, $\delta_2$ and $\delta_3$ as we take the limit $a\rightarrow +\infty$ \cite{toms:1980:2805, chung:1998} . This is done by making use of the conditions \eqref{ren1}, \eqref{ren2} and \eqref{ren3} taken at $\Phi=0$. The condition \eqref{ren1} fix the renormalization constant $\delta_1$, i.e.,
\begin{equation}
 \label{lambda_const_DN_t}
 \begin{aligned}
 \frac{\delta_1}{4!} &= \frac{\lambda^2 \text{ln}\left(\frac{\mu^2}{m^2}\right)}{256\pi^2\sqrt{1+\chi}} \ .
 \end{aligned}
\end{equation}
Furthermore, the renormalization conditions \eqref{ren2} and \eqref{ren3} fix, respectively, the constants $\delta_2$ and $\delta_3$, providing 
\begin{equation}
 \label{m2_const_DN_t}
 \begin{aligned}
 \frac{\delta_2}{2} = \frac{\lambda m^2}{64\pi^2\sqrt{1+\chi}} + \frac{\lambda m^2 \text{ln}\left(\frac{\mu^2}{m^2}\right)} {64\pi^2\sqrt{1+\chi}} \ ,
 \end{aligned}
\end{equation}
and
\begin{equation}
 \label{m4_const_DN_t}
 \begin{aligned}
 \delta_3= \frac{3m^4}{128\pi^2\sqrt{1+\chi}} + \frac{m^4\text{ln}\left(\frac{\mu^2}{m^2}\right)}{64\pi^2\sqrt{1+\chi}} \ .
 \end{aligned}
\end{equation}

The renormalization constants above when taking into account in Eq. \eqref{effec_potent_DN_t} allow us to obtain the renormalized effective potential at one-loop level
\begin{equation}
 \label{ren_effec_potent_DN_t}
 \begin{aligned}
  V_{\text{eff}}^{\text{R}}(\Phi) &= \frac{m^2\Phi^2}{2} + \frac{\lambda \Phi^4}{4!} - \frac{3b^4}{128\pi^2\sqrt{1+\chi}} + \frac{3m^4}{128\pi^2\sqrt{1+\chi}} + \frac{b^3}{24\pi \sqrt{1+\chi} a} \\ & \ \ + \frac{m^2\lambda \Phi^2}{64\pi^2\sqrt{1+\chi}} + \frac{b^4\text{ln}\left(\frac{b^2}{m^2}\right) }{64\pi^2 \sqrt{1+\chi}}- \frac{b^2}{8\pi^2\sqrt{1+\chi}a^2}\sum_{n=1}^{\infty}\frac{K_2\left(2ban\right)}{n^2} \ .
 \end{aligned}
\end{equation}
This expression for the renormalized effective potential is clearly affected by the Lorentz symmetry violation parameter $\chi$, as it should be.

The renormalized effective potential, \eqref{ren_effec_potent_DN_t}, when taken at the vacuum state $\Phi=0$, provides a non-vanishing vacuum Casimir-like potential energy by unit area of the plates, given by,
\begin{equation}
 \label{dens_ener_cas_exa_DN_t}
 \frac{{E}_{C}}{L^2} = a V_{\text{eff}}^{\text{R}}(0) = -\frac{m^2}{8\pi^2\sqrt{1+\chi}a}\sum_{n=1}^{\infty}\frac{K_2\left(2amn\right)}{n^2} \ .
\end{equation}

As we can see the Casimir potential energy density above is affected by the Lorentz symmetry violation parameter $\chi$ through a multiplicative factor. Although this potential energy is given in terms of a sum of the modified Bessel functions, $K_{2}(2man)$, it is a convergent expression, since for large values of $n$ this function is exponentially suppressed. 

It is possible to provide a closed expressions for the asymptotic behaviors as $am\gg 1$ and $am\ll 1$, of the vacuum potential energy density.  Thus, in the case $am\gg 1$, by using the asymptotic expression for the modified Bessel function for large argument \cite{grad}, we get 
 \begin{equation} \label{cas_energ_DN_t_large}
  \frac{E_{C}}{L^2} \approx - \frac{1}{16\sqrt{1+\chi}} \left(\frac{m}{\pi a}\right)^{3/2} e^{-2am} \ .
 \end{equation}
In this limit, the dominant term in \eqref{dens_ener_cas_exa_DN_t} is for $n=1$, and we can clearly see that for large values of $am$ the vacuum potential energy density is exponentially suppressed.

As to the case when $am \ll 1$  it is convenient first to use the integral representation for the modified Bessel function \cite{abramowitz:1948}:
 \begin{equation}
 \label{int_bessel_form}
 K_{\nu}(z)=\frac{\sqrt{\pi}\left(\frac{1}{2}z\right)^{\nu}}{\Gamma(\nu+1/2)}\int_1^{\infty}e^{-zt}\left(t^2-1\right)^{\nu-1/2} dt \  .
 \end{equation}
By substituting the above representation into \eqref{dens_ener_cas_exa_DN_t}, it is possible to develop the sum over $n$. Doing this we get:
 \begin{eqnarray}
 \label{potential}
 \frac{E_C}{L^2}=-\frac{am^4}{6\pi^2\sqrt{1+\chi}}\int_1^\infty dv\frac{(v^2-1)^{3/2}}{e^{2amv}-1}  \  .
 \end{eqnarray}
 In the regime of $am<<1$, the integral in \eqref{potential} is dominated by large values of $v$; so we may approximated \eqref{potential}  to
  \begin{eqnarray}
 \label{potential1}
 \frac{E_C}{L^2}\approx-\frac{am^4}{6\pi^2\sqrt{1+\chi}}\int_1^\infty dv\frac{v^3}{e^{2amv}-1}  \  .
 \end{eqnarray}
Now we can obtain an expression to the integral in \eqref{potential1}, which developing a series expansion in powers of $am<<1$, provides,
\begin{equation}
  \frac{E_{C}}{L^2} \approx -\frac{\pi^2}{1440 \sqrt{1+\chi} a^3}+\frac{m^3}{36 \pi^2 \sqrt{1+\chi}} - \frac{a m^4}{48 \pi^2 \sqrt{1+\chi}} \ .
  \label{asy2}
 \end{equation}
Note that the leading term in the first term on the r.h.s of \eqref{asy2} is the contribution of the massless scalar field, which becomes an exact expression in the limit $m\rightarrow 0$.
Moreover, we also recover from Eq. \eqref{dens_ener_cas_exa_DN_t} results for the Casimir effect for a real scalar field which satisfies Dirichelet and Neumann boundary conditions on two parallel plates without Lorentz violation \cite{bordag:2001,milton:2003, cruz:2017}. 

As it has been said before, the two-loop contribution to the effective potential comes from the two graphs to the effective Euclidian action given in \cite{toms:1980:2805, toms:1980:334357}. As we are only interested in the two-loop contribution to the vacuum energy, its only nonzero term, at $\Phi=0$, is given by 
\begin{equation} 
\label{vac_two_loop_contr_potent}
 V^{(2)}(\Phi=0) = \frac{\lambda}{8} S_1(\Phi=0) \ .
\end{equation}
The function $S_1(\Phi)$ is obtained by means of the expression
\begin{equation} 
\label{funct_S1_DN_t}
 S_1(\Phi) = \left \{ \sum_{n=1}^{\infty} \frac{1}{a} \int \frac{d^3k}{(2\pi)^3}\left[(1+\chi)k_t^2+k^2+\frac{\pi^2 n^2}{a^2}+m^2+\frac{\lambda \Phi^2}{2}\right]^{-s}\right \}^{2} \ ,
\end{equation}
where $s$ has been introduced in order to regularize the expression above. After we subtract the divergente part of \eqref{funct_S1_DN_t} one should take $s=1$. This allows us to write the finite contribution of the function $ S_1(\Phi)$ in terms of the zeta function \eqref{zet_funct_DN_t_2} as
\begin{equation}
 \label{funct_S1_DN_t_f2}
 S_1(\Phi) = \left[\frac{\zeta_R(1)}{V_3 a}\right]^2 \ .
\end{equation}
Note that $\zeta_R(1)$ is the zeta function \eqref{zet_funct_DN_t_2} taken at $s=1$ after subtracting the divergent part of it given by the second term on the r.h.s. This term, when divided by $V_3 a$, is independent of $a$ and presents a divergent contribution at $s=1$ proportional to $\frac{\Gamma(s-2)}{\Gamma(s)}\approx \frac{s}{1-s}$. As it is usually done, this term should be subtracted since it does not depend on the boundary condition parameter, that is, $a$. Hence, Eqs. \eqref{vac_two_loop_contr_potent} and \eqref{funct_S1_DN_t_f2} provide the two-loop contribution to the effective potential as 
\begin{equation} 
 \label{2_loop_cas_energ_DN_t}
 \frac{E^{(\lambda)}_C}{L^2} = aV^{(2)}(0) = \frac{m^2 \lambda}{128 \pi^4 (1+\chi)a} \left[\sum_{n=1}^{\infty}\frac{K_1\left(2amn\right)}{n}\right]^2 \ .
\end{equation}
Also we can see that $\frac{E^{(\lambda)}_C}{L^2}$ depends on the Lorentz violation parameter by a multiplicative factor; moreover, it is an exact and a convergent function, which can be seen by noting that the modified Bessel function $K_{1}(2man)$ is exponentially suppressed for large values of $n$.
\begin{figure}[h!]
	\centering
	\subfloat{{\includegraphics[width=17cm]{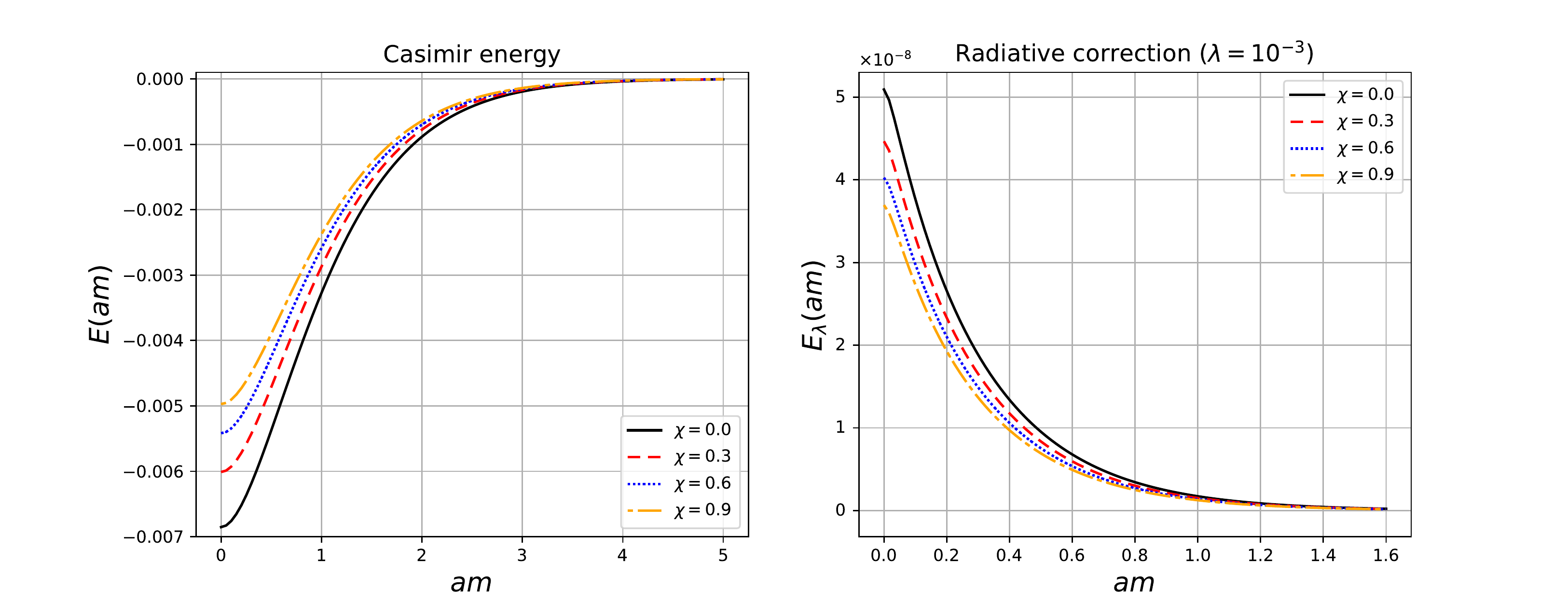} }}%
	\qquad
	\caption{The behaviors of the Casimir energy per unity area $E(am)=\frac{a^3}{L^2}E_C$ given by \eqref{dens_ener_cas_exa_DN_t} as function of $am$ is exhibited in the left panel, and  the two-loop contribution to the Casimir energy per unit area $E_{\lambda}(am)=\frac{a^3}{L^2}E^{(\lambda)}_C$ in Eq. \eqref{2_loop_cas_energ_DN_t} as function of $am$, is presented in the right panel considering $\lambda=10^{-3}$. For both plots different values for the parameter $\chi$, is considered.}
	\label{figure2}%
\end{figure}

We can also obtain a closed expressions for \eqref{2_loop_cas_energ_DN_t} in the regimes when $am \gg 1$ and $am \ll 1$. Considering first the case $am>>1$,  we have
\begin{equation}
\label{app1}
\frac{E^{(\lambda)}_C}{L^2} \approx \frac{\lambda  me^{-4 a m}}{512 \pi^3 (1+\chi) a^2} \ ,
\end{equation}
which is dominated by the term $n=1$ in the sum and it is exponentially suppressed. This feature is shown in Fig.\ref{figure2}. 

For the opposite regime, that is when $am \ll 1$, we should use again the integral representation \eqref{int_bessel_form} for the modified Bessel function $K_{\mu}(x)$. Developing a similar procedure as before, we obtain
\begin{equation}
\begin{aligned}
\frac{E^{(\lambda)}_C}{L^2} & \approx \frac{\lambda }{18432 (1+\chi)a^3} - \frac{\lambda  m}{768 \pi^2(1+\chi)a^2} \ ,
\label{expansion}
\end{aligned}
\end{equation}
which give us as the leading contribution the massless scalar field expression in the first term on the r.h.s of it. This becomes an exact expression in the limit $m\rightarrow 0$.

In the left panel of Fig.\ref{figure2}, we exhibit the behavior of the Casimir energy per unity area given by Eq. \eqref{dens_ener_cas_exa_DN_t}, as function of the dimensionless parameter $ma$, considering different values for the parameter $\chi$. We can see that it increases as the Lorentz symmetry violation parameter increases. In the right plot, on the other hand, we exhibit the two-loop correction to the Casimir energy per unity area, Eq. \eqref{2_loop_cas_energ_DN_t},  as function of $am$. It is clear that both plots are in agreement with the asymptotic behaviors \eqref{cas_energ_DN_t_large} and \eqref{asy2} for the Casimir energy, and with \eqref{app1} and \eqref{expansion} for the two-loop correction.
\begin{figure}[h!]
    \centering
    \includegraphics[scale=0.5]{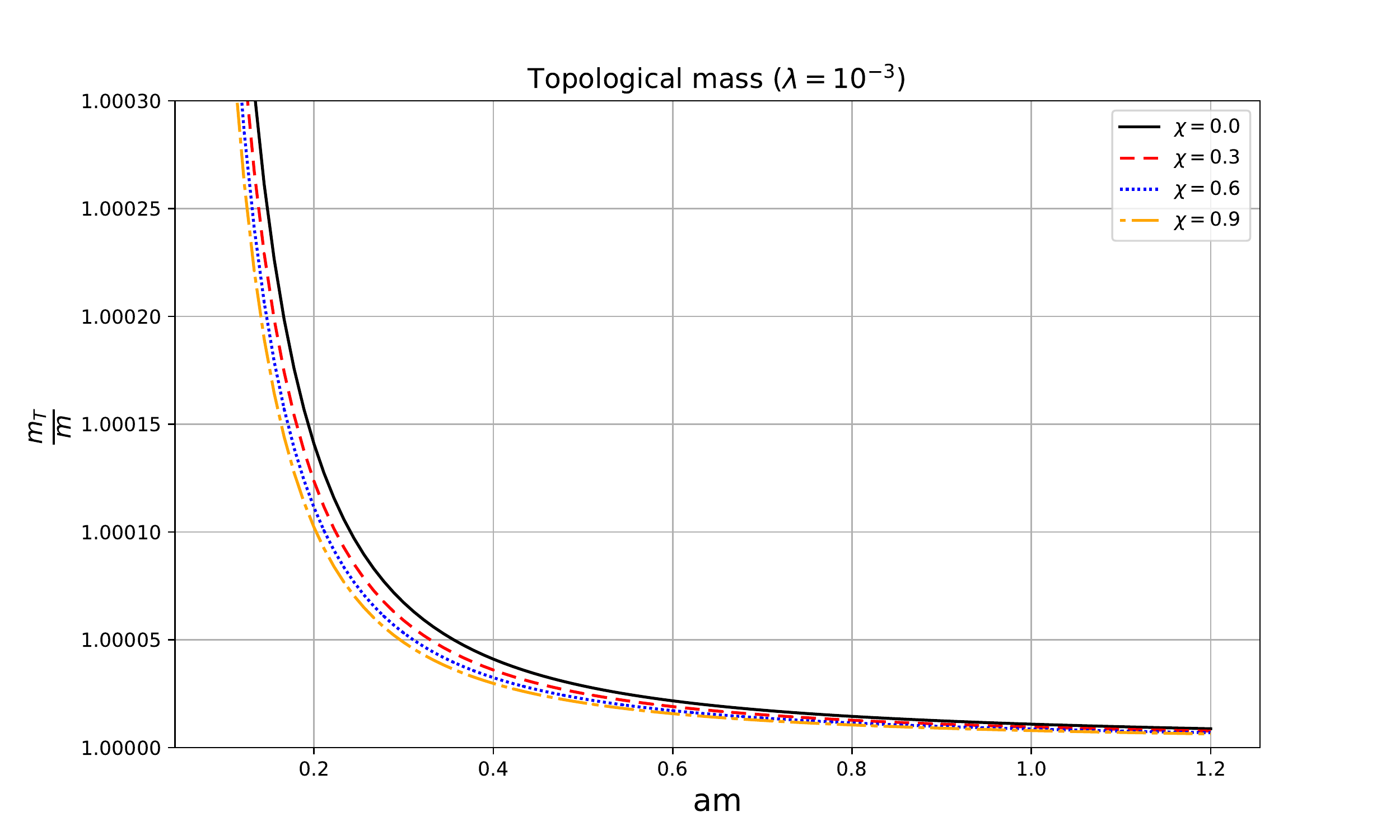}
    \caption{The behavior of the ration of the topological mass by the scalar field mass as function of $am$ is plotted assuming $\lambda=10^{-3}$ and different values for the Lorentz violating parameter $\chi$.}
    \label{figure3}
\end{figure}

At one-loop level the massive scalar field we are considering will get quantum corrections to its mass. This correction can be obtained by using condition \eqref{ren2}, at $\Phi=0$, for the renormalized effective potential \eqref{ren_effec_potent_DN_t} at one-loop level, i.e.,
\begin{eqnarray} 
\label{topological_mass_exata}
 m_{\text{T}}^2 &=& \frac{d^2V_{\text{eff}}^{\text{R}}(\Phi)}{d\Phi^2}\bigg{|}_{\Phi=0}\nonumber\\
&=& m^2 \left[1 + \frac{\lambda}{16 \pi \sqrt{1+\chi} a m} + \frac{\lambda}{8 \pi^2 \sqrt{1+\chi} am} \sum_{n=1}^{\infty} \frac{K_1\left(2amn\right)}{n} \right] \ .
\end{eqnarray}
This expression presents a topological contribution, which depends on $a$, to the mass $m$ of the scalar field given by the terms proportional to the self-coupling constant $\lambda$. Even for the massless scalar field, which remains massless at the tree-level, there appears a topological mass  generated at one-loop correction  as we can see from Eq. \eqref{topological_mass_exata}. Moreover, the third term inside the bracket is convergent, since it is exponentially suppressed for large values of $n$. In fact, the leading contribution for $am\gg 1$ is given by the first term on the r.h.s of \eqref{topological_mass_exata}. This asymptotic behavior can be obtained by using the expressions for the modified Bessel function for large argument \cite{grad}. The asymptotic expression for \eqref{topological_mass_exata} in the regime  $am\gg 1$ is: 
\begin{equation}
m_{\text{T}}^2 \approx m^2 +  \frac{\lambda m}{16 \pi \sqrt{1+\chi} a} + \frac{\lambda}{16 \pi ^{3/2}} \sqrt{\frac{m}{a^3}} \frac{e^{-2 a m}}{ \sqrt{1+\chi}} \ .
\label{asm1}
\end{equation}
On the other hand, for $am\ll 1$, the leading term is mass independent, followed by terms that depend on the mass of the scalar field. Once more this behavior can be obtained by using the integral representation for the function $K_\mu(z)$, \eqref{int_bessel_form}. After some intermediate steps we obtain:
	\begin{equation}
	\begin{aligned}
	m_{\text{T}}^2 & \approx  \frac{\lambda}{96 \sqrt{1+\chi} a^2} + \frac{\lambda m}{16 \pi  \sqrt{1+\chi} a} - \frac{\lambda m}{8 \pi^2 \sqrt{1+\chi} a} +m^2 \ .
	\label{asm2}
	\end{aligned}
	\end{equation}
The asymptotic results, Eqs. \eqref{asm1} and \eqref{asm2},  are in agreement with the plot of  Fig.\ref{figure3} which exhibits the behavior of the ratio of the topological mass to the field's mass itself, $m_T/m$, as function of $am$ for different parameter $\chi$.  Note that the topological mass decreases as $\chi$ increases. 
%

\subsubsection{Spacelike vector}
%
Now we consider the case that constant 4-vector is space-like, that is, of the types $u^x=(0,1,0,0)$, $u^y=(0,0,1,0)$ and $u^z=(0,0,0,1)$. The first two types are parallel to the plates and provide essentially the same results while the third type is perpendicular to the plates. Let us then begin by considering the parallel cases by assuming, for instance, that a broke of Lorentz symmetry happens in the $x$-direction, i.e.,  $u^x = (0,1,0,0)$. This give us the set of eigenvalues:
\begin{equation}
 \Lambda_{\beta} = k^2 + (1-\chi)k_x^2 + \frac{\pi^2n^2}{a^2} + m^2 + \frac{\lambda}{2}\Phi^2 \ ,
\end{equation}
where $k^2=k_t^2+k_y^2$. Thereby, from Eq. \eqref{operator}, the set of eigenvalues above is associated with the elliptic-operator given by
\begin{equation}
 \Delta = -\partial_{\mu}\partial^{\mu} + \chi \partial_x^2 + m^2 + \frac{\lambda}{2}\Phi^2 \ .
\end{equation}
So, in this case, the zeta function \eqref{zet_funct_DN_t_0} is written as
\begin{equation}
 \zeta(s) = \frac{V_3}{(2\pi)^3} \int d^3k \sum_{n=1}^{\infty}\left[k^2 + (1-\chi)k_x^2 + \frac{\pi^2n^2}{a^2} + m^2 + \frac{\lambda}{2}\Phi^2\right]^{-s} \ .
\end{equation}
The integrals in $k_t$, $k_x$ and $k_y$ can be solved using the identity \eqref{identity}, providing 
\begin{equation}
 \zeta(s) = \frac{V_3}{(2\pi)^3}\frac{\pi^{3/2}}{\sqrt{1-\chi}}\frac{\Gamma(s-3/2)}{\Gamma(s)}w^{3-2s}\sum_{n=1}^{\infty}\left(n^2+\nu^2\right)^{3/2-s} \ ,
 \label{zetatl1}
\end{equation}
where
\begin{equation}
 \nu^2 = \frac{\lambda \Phi^2}{2w^2}+\frac{m^2}{w^2} \ \ \ \ \text{and} \ \ \ \ w=\frac{\pi}{a} \ .
\end{equation}
Clearly, we can see that this result is very similar to the one obtained in the timelike case, with a different dependence on the Lorentz violation parameter, $\chi$. In fact, the timelike and spacelike expressions for the zeta function \eqref{zet_funct_DN_t_1} and \eqref{zetatl1}, respectively, are related by the change $\chi \rightarrow -\chi$. Hence, we expect that all the results for the vacuum energy, its loop-correction and the topological mass will be related by this same change.

Let us now turn to the most important type of the spacelike Lorentz symmetry violation, namely, the one in the perpendicular direction to the plates given by
\begin{equation}
 u^{z} = (0,0,0,1) \ .
\end{equation}
This provides, from Eq. \eqref{operator}, the following differential elliptic operator:
\begin{equation}
 \label{diff_operador_DN_z}
 \Delta = - \partial_{\mu} \partial^{\mu} + \chi \partial_z^2 + m^2 + \frac{\lambda \Phi^2}{2} \ .
\end{equation}
Consequently, the set of eigenvalues associated with this operator is found to be
\begin{equation}
\label{eigenvalue_DN_z}
 \Lambda_{\beta} = k^2 + \left(1-\chi \right)\frac{\pi^2 n^2}{a^2} + m^2 + \frac{\lambda \Phi^2}{2} \ .
\end{equation}
Note that $k^2=k_t^2+k_x^2+k_y^2$. Thereby, the zeta function \eqref{zet_funct_DN_t_0}, taking into consideration \eqref{eigenvalue_DN_z}, is written as
\begin{equation} 
\label{zeta_funct_s_DN_z}
 \zeta(s) = \frac{V_3}{(2 \pi)^3} \sum_{n=1}^{\infty} \int d^3k \left[ k^2 + \left(1-\chi \right) \frac{\pi^2 n^2}{a^2} + m^2 + \frac{\lambda \Phi^2}{2}\right]^{-s} \ ,
\end{equation}
where, again, $V_3$ is the continuum volume associated with the dimensions $t,x,y$ and $d^3k=dk_tdk_xdk_y$. Thus, the identity \eqref{identity} allows us to perform the integrals in $k_t$, $k_x$ and $k_y$ in Eq. \eqref{zeta_funct_s_DN_z}. The resulting expression, in analogy with Eq. \eqref{zetatl1},  can be written in terms of the Epstein-Hurwitz function \eqref{EH_funct}, providing that
\begin{equation}
 \begin{aligned}
  \zeta(s) = \frac{V_3}{(2\pi)^3}  \bigg{[} & -\frac{\pi^{3/2} w^{-2\bar{s}} \nu^{-2\bar{s}}}{2} \frac{\Gamma(\bar{s})}{\Gamma(\bar{s}+3/2)} + \frac{\pi^{1/2} \pi^{3/2} w^{-2\bar{s}}}{2}\frac{\Gamma(\bar{s}-1/2)}{\Gamma(\bar{s}+3/2)}\nu^{1-2\bar{s}} \\ & + \frac{2^{1-\bar{s}}(2\pi)^{2\bar{s}-1/2} \pi^{3/2} w^{-2\bar{s}}}{\Gamma(\bar{s}+3/2)} \sum_{n=1}^{\infty} n^{2\bar{s}-1} f_{(\bar{s}-1/2)}(2\pi n \nu) \bigg{]} \ ,
 \end{aligned}
 \label{zetaSL}
\end{equation}
where $\bar{s}=s-3/2$ and 
\begin{equation}
 \begin{aligned}
  \nu^2 = \frac{\lambda \Phi^2}{2w^2} + \frac{m^2}{w^2} \ \ \ \ \ \text{and} \ \ \ \ \ w \equiv \sqrt{1-\chi} \frac{\pi}{a} \ .
 \end{aligned}
\end{equation}
Note that one should consider the limit $s\rightarrow 0$, or analogously $\bar{s}\rightarrow -3/2$. In this limit, we can get the expressions for $\zeta(0)$ and $\zeta^{\prime}(0)$, respectively, given by
\begin{equation}
\label{zeta_s0_DN_z}
 \zeta(0) = \frac{V_3 a}{2(2\pi)^4} \frac{\pi^2 b^4}{\sqrt{1-\chi}} \ ,
\end{equation}
\begin{equation}
\label{zeta_prim_s0_DN_z}
 \begin{aligned}
 \zeta^{\prime}(0) = \frac{V_3 L}{(2\pi)^3} \left[ - \frac{2\pi^{2}}{3} \frac{b^3}{a} +\frac{3}{8}\frac{\pi b^4}{\sqrt{1-\chi}} - \frac{\pi}{2}\frac{b^4 \text{ln}(b)}{\sqrt{1-\chi}} + \frac{2\pi \sqrt{1-\chi} b^2}{a^2} \sum_{n=1}^{\infty} \frac{K_2\left(\frac{2abn}{\sqrt{1-\chi}}\right)}{n^2} \right] \ .
 \end{aligned}
\end{equation}
Consequently, from Eq. \eqref{one_loop_potential_DN_t}, the one-loop correction to the effective potential is 
\begin{equation}
\label{cont_potent_1loop_DN_z}
 \begin{aligned}
  V^{(1)}(\Phi) & = \frac{b^3}{24 \pi  a} - \frac{3 b^4}{128 \pi^2 \sqrt{1-\chi}} + \frac{b^4 \text{ln}\left(\frac{b^2}{\mu^2}\right)}{64 \pi^2 \sqrt{1-\chi}} \\ & \ \ \ - \frac{b^2 \sqrt{1-\chi}}{8 \pi^2 a^2} \sum_{n=1}^{\infty}\frac{K_2\left(\frac{2abn}{\sqrt{1-\chi}}\right)}{n^2} \ ,
 \end{aligned}
\end{equation}
where the parameter $b$ is defined as
\begin{equation}
 b = \sqrt{\frac{\lambda \Phi^2}{2} + m^2} \ .
\end{equation}
Hence, the effective potential up to one-loop correction, given by the expression \eqref{effP}, is obtained as
\begin{equation} \label{eff_potential_DN_z}
 \begin{aligned}
  V_{\text{eff}}(\Phi) & = \frac{m^2\Phi^2}{2} + \frac{\lambda \Phi^4}{4!} + \frac{\Phi^2}{2}\delta_2 + \frac{\Phi^4}{4!}\delta_1 + \delta_3 + \frac{b^3}{24 \pi  a} - \frac{3 b^4}{128 \pi^2 \sqrt{1-\chi}} \\ & \ \ \ + \frac{b^4 \text{ln}\left(\frac{b^2}{\mu^2}\right)}{ 64\pi^2 \sqrt{1-\chi}} - \frac{b^2 \sqrt{1-\chi}}{8 \pi^2 a^2} \sum_{n=1}^{\infty}\frac{K_2\left(\frac{2abn}{\sqrt{1-\chi}}\right)}{n^2} \ .
 \end{aligned}
\end{equation}
We should now obtain the renormalization constants $\delta_1$, $\delta_2$ and $\delta_3$, which can be done by using the conditions \eqref{ren1}, \eqref{ren2} and \eqref{ren3}, respectively. Thus, the renormalization constant $\delta_1$ is found to be 
\begin{equation}
\label{lambda_const_DN_z}
 \frac{\delta_1}{4!} = \frac{\lambda^2 \text{ln}\left(\frac{\mu^2}{m^2}\right)}{256\pi^2\sqrt{1-\chi}} \ , 
\end{equation}
\begin{equation}
\label{m2_const_DN_z}
 \frac{\delta_2}{2!} = \frac{m^2\lambda}{64\pi^2\sqrt{1-\chi}} + \frac{m^2\lambda \text{ln}\left(\frac{\mu^2}{m^2}\right)}{64\pi^2\sqrt{1-\chi}},
\end{equation}
and
\begin{equation}
\label{m4_const_DN_z}
 \delta_3 = \frac{3m^4}{128\pi^2\sqrt{1-\chi}} + \frac{m^4 \text{ln}\big(\frac{\mu^2}{m^2}\big)}{64\pi^2\sqrt{1-\chi}}.
\end{equation}
Hence, substituting the renormalization constants above in Eq. \eqref{eff_potential_DN_z}, the renormalized effective potential, up to one-loop correction, is found to be
\begin{equation}
\label{ren_eff_potential_DN_z}
 \begin{aligned}
  V_{\text{eff}}^{\text{R}}(\Phi) &= \frac{m^2\Phi^2}{2} + \frac{\lambda \Phi^4}{4!} + \frac{b^3}{24\pi a} - \frac{3b^4}{128\pi^2\sqrt{1-\chi}} + \frac{m^2 \lambda \Phi^2}{64 \pi^2 \sqrt{1-\chi}} + \frac{3m^4}{128 \pi^2 \sqrt{1-\chi}} \\ & + \frac{b^4}{64 \pi^2 \sqrt{1-\chi}}\text{ln}\left(\frac{b^2}{m^2}\right) - \frac{b^2 \sqrt{1-\chi}}{8 \pi^2 a^2} \sum_{n=1}^{\infty}\frac{K_2\left(\frac{2abn}{\sqrt{1-\chi}}\right)}{n^2} \ .
 \end{aligned}
\end{equation}

The vacuum energy per unit area of the plates is obtained when we take the vacuum state ($\Phi=0$). Thus, from Eq. \eqref{ren_eff_potential_DN_z} we get
\begin{equation}
\label{1loop_cas_energ_DN_z}
  \frac{E_{\text{C}}}{L^2} = a V_{\text{eff}}^{R}(0) = - \frac{m^2 \sqrt{1-\chi}}{8 \pi^2 a} \sum_{n=1}^{\infty}\frac{K_2\left(\frac{2 a m n}{\sqrt{1-\chi}}\right)}{n^2} \ ,
\end{equation}
which is convergent and, therefore, finite. This expression for the vacuum energy per unit area is exponentially suppressed for $ma\gg 1$ and provides the massless scalar field expression for $ma\ll1$.

We can mathematically obtain asymptotic expression for \eqref{1loop_cas_energ_DN_z} in the regimes $ma\ll 1$ and $ma\gg1$. By considering the latter we have 

 \begin{equation}
 \label{app2}
   \frac{E_C}{L^2} \approx -\frac{\sqrt{1-\chi}}{16}\left(\frac{m}{\pi a}\right)^{3/2} e^{-\frac{2am}{\sqrt{1-\chi}}} \ ,
 \end{equation}
which, as mentioned before, is exponentially suppressed. It is the dominant term for $n=1$ in the sum. 

As to the limit $am \ll 1$, we should first use the integral representation for the modified Bessel function. $K_{\mu}(x)$, in Eq. \eqref{int_bessel_form}. This provides
 \begin{equation}
  \frac{E_C}{L^2} \approx -\frac{\pi^2 (1-\chi)^{3/2}}{1440 a^3} + \frac{m^3}{36 \pi^2} - \frac{a m^4}{48 \pi^2
  \sqrt{1-\chi }} \ .
  \label{enerper}
 \end{equation}
 Note that the dominant term is the first one on the r.h.s and represents the vacuum energy per unit area in the massless scalar field case.  

Let us turn to the two-loop contribution to the effective potential calculated at $\Phi=0$ in the case the $4$-vector is orthogonal to the parallel plates. The $S_1(\Phi)$ function in Eq. \eqref{vac_two_loop_contr_potent} is now written as
\begin{equation}
\label{funct_S1_DN_z}
 S_1(\Phi) = \left \{ \sum_{n=1}^{\infty} \frac{1}{a} \int \frac{d^3k}{(2\pi)^3} \left [ k^2 + (1-\chi)\frac{\pi^2 n^2}{a^2} + m^2 + \frac{\lambda \Phi^2}{2} \right ]^{-s} \right \}^2 \ ,
\end{equation}
which should be taken at $s=1$ after subtracting the divergent contribution. This can be done by using the zeta function \eqref{zetaSL} in a similar way as in the previous case shown in Eq. \eqref{funct_S1_DN_t_f2}. In the present case, the divergent contribution comes from the second term on the r.h.s of Eq. \eqref{zetaSL} and, after subtracted, the two-loop contribution $V^{(2)}(0)$ at $s=1$ can be found, leading to the two-loop correction to the vacuum energy 
\begin{equation}
 \label{2_loop_cas_energ_DN_z}
 \frac{E^{(\lambda)}_C}{L^2} = a V^{(2)}(0) = \frac{m^2 \lambda}{128 \pi^4 a} \left[\sum_{n=1}^{\infty} \frac{K_1\left(\frac{2amn}{\sqrt{1-\chi}}\right)}{n}\right]^2 \ .
\end{equation}
This is a completely convergent expression since the modified Bessel function, $K_{\mu}(x)$, is exponentially suppressed. This is also clear when one consider the asymptotic limit $ma\gg 1$. In the opposite limit, $ma\ll 1$, the expression for the two-loop contribution to the vacuum energy for a massless scalar field is obtained, as leading contribution. 

\begin{figure}[h!]
	\centering
	\subfloat{{\includegraphics[width=17cm]{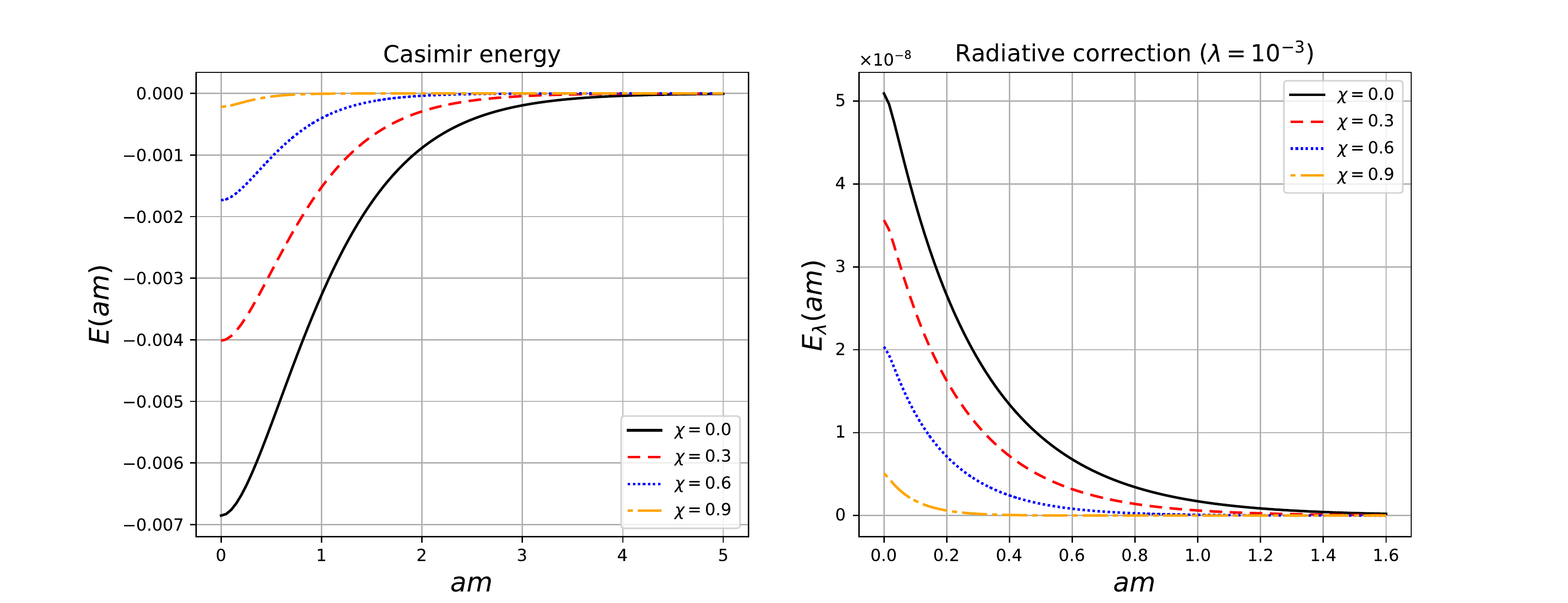} }}
	\qquad
	\caption{The behavior of the Casimir energy per unity area $E(am)=\frac{a^3}{L^2}E_C$ given by Eq. \eqref{1loop_cas_energ_DN_z} as function of $am$ is presented in the left panel, and the two-loop correction $E_{\lambda}(am)=\frac{a^3}{L^2}E^{(\lambda)}_C$ , given by \eqref{2_loop_cas_energ_DN_z}, as function of $am$ is in the right one. In the latter we have taken $\lambda=10^{-3}$ and considered different values for $\chi$.}
	\label{figure4}
\end{figure}

We can also obtain the asymptotic expressions for \ref{2_loop_cas_energ_DN_z} in the regimes $ma\ll1$ and $ma\gg 1$. In the latter, we have
 \begin{equation}
  \frac{E^{(\lambda)}_C}{L^2} \approx \frac{\lambda  m \sqrt{1-\chi} e^{-\frac{4 a m}{\sqrt{1-\chi}}}}{512 \pi^3 a^2} \ ,
 \end{equation}
which is exponentially suppressed and dominated by the term $n=1$ in the sum. 

In the opposite limit, $ma\ll 1$, we need to use the integral representation \eqref{int_bessel_form} for the modified Bessel function, $K_{\mu}(x)$. This provides 
 \begin{equation}
  \begin{aligned}
  \frac{E^{(\lambda)}_C}{L^2} & \approx \frac{\lambda (1-\chi)}{18432 a^3} - \frac{\lambda  m \sqrt{1-\chi}}{768 \pi^2 a^2}  \ .
  \label{SLVE}
  \end{aligned}
 \end{equation}
Note that the first term on the r.h.s is the massless scalar field contribution which becomes exact when $ma\rightarrow 0$. 

In the left panel of Fig.\ref{figure4} we plot the behavior of the Casimir energy per unity area, \eqref{1loop_cas_energ_DN_z}, as function of $am$ considering different values for the parameter $\chi$. This plot shows that as $\chi$ increases, the vacuum energy also increases. On the other hand, in the right figure we plot the two-loop correction to the vacuum energy per unity area, \eqref{2_loop_cas_energ_DN_z}, as function of $am$, assuming $\lambda=10^{-3}$. This plot shows that as $\chi$ increases, the two-loop correction decreases. Moreover, one should note an importante feature here, namely, the vacuum energy \eqref{1loop_cas_energ_DN_z} and its radiative correction \eqref{2_loop_cas_energ_DN_z} not only depends on, $\chi$, by means of a multiplicative factor but also depends on, $\chi$, in the argument of the modified Bessel function, $K_{\mu}(x)$. This stronger dependence is shown in the plots of Fig.\ref{figure4}. The shift in the curves are stronger than in the previous timelike case. 

A topological mass in this case will also be generated and can be obtained by using the condition \eqref{ren2}. Thus, by applying the latter in the renormalized effective potential \eqref{ren_eff_potential_DN_z} we find
\begin{equation}
 \label{topol_mass_DN_z}
 m_{\text{T}}^2 = m^2 \left[1 + \frac{\lambda}{16 \pi a m} + \frac{\lambda}{8 \pi^2 a m} \sum_{n=1}^{\infty} \frac{K_1\left(\frac{2 a m n}{\sqrt{1-\chi}}\right)}{n} \right] \ .
\end{equation}
This is an exact convergent expression and in the limit $ma\gg 1$ is dominated by the first term on the r.h.s while in the limit $ma\ll 1$ is dominated by the third term, the massless scalar field contribution. 
\begin{figure}[h!]
    \centering
    \includegraphics[scale=0.5]{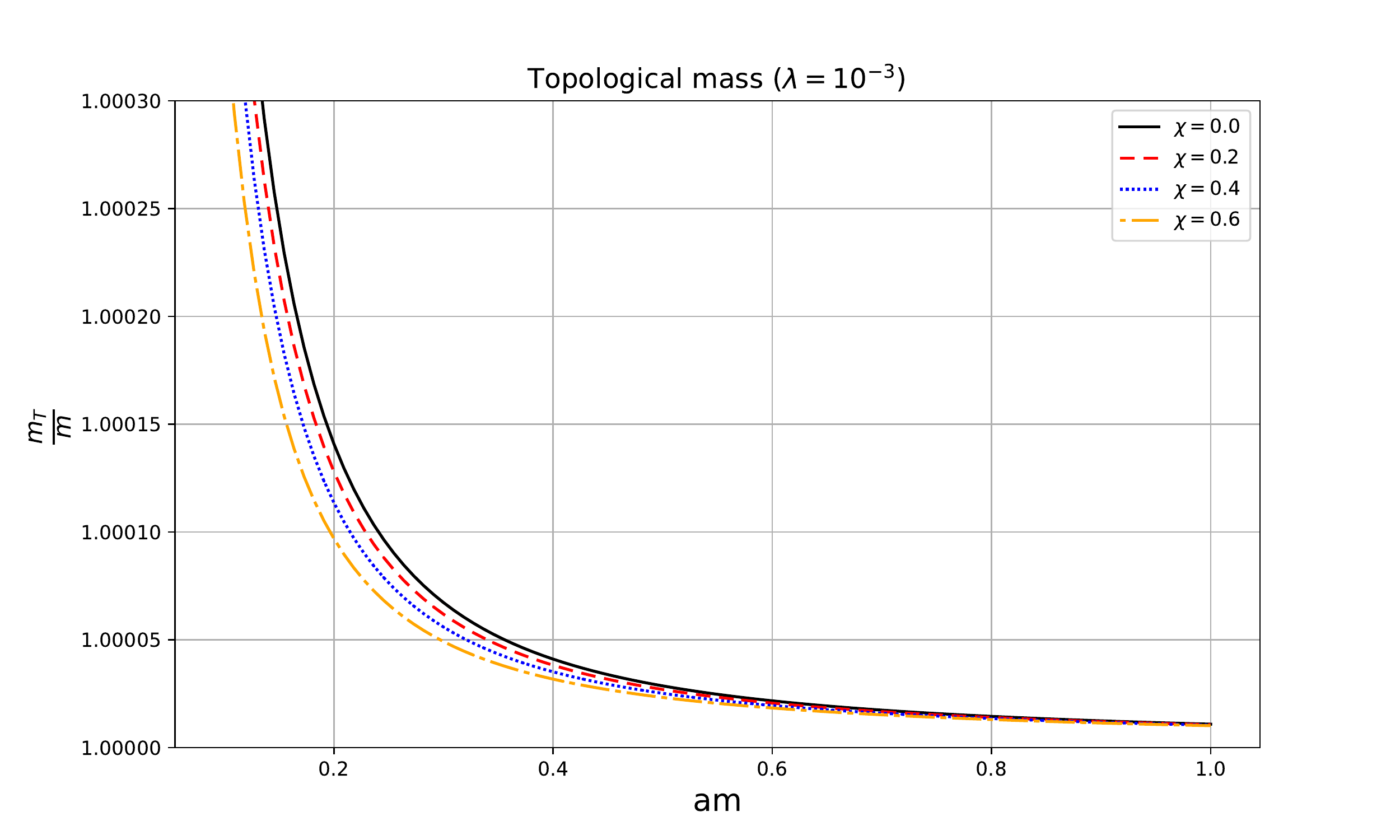}
    \caption{Plot exhibiting the behavior of the ration of the topological mass by the mass of the field as function of $am$. In the plot is consider 
    $\lambda=10^{-3}$ and different variation of $\chi$.}
    \label{figure5}
\end{figure}

Mathematically, the asymptotic behaviour $ma\gg 1$ of \eqref{topol_mass_DN_z} is given by 
 \begin{equation}
 \label{top_mas1}
  m^2_{\text{T}} \approx m^2 + \frac{\lambda m}{16 \pi  a} + \frac{\lambda \sqrt[4]{1-\chi}}{16 \pi^{3/2}} \sqrt{\frac{m}{a^3}}  
   e^{-\frac{2 a m}{\sqrt{1-\chi}}} \ ,
 \end{equation}
while the asymptotic behaviour $ma\ll 1$ is
 \begin{equation}
 \label{top_mas2}
  m^2_{\text{T}} \approx \frac{\lambda \sqrt{1-\chi}}{96 a^2} + \frac{\lambda m}{16 \pi a} +m^2   \ . 
 \end{equation}
 The asymptotic results, Eqs. \eqref{top_mas1} and \eqref{top_mas2}, are in agreement with the plot of Fig.\ref{figure5} which
 exhibits the behavior of the ratio of the topological mass to the field's mass as function of $ma$ for a fixed value of $\lambda$
 and different values of $\chi$. Note that, due to the dependence of the topological mass \eqref{topol_mass_DN_z} on $\chi$ in the argument of the modified Bessel function, $K_{\mu}(x)$, the curves in Fig.\ref{figure5} are shifted down, as $\chi$ increses, more than in the timelike case. 

It is important to point out that all the results obtained up to here, adopting Dirichlet boundary condition, for the timelike and spacelike types of Lorentz symmetry violation are the same ones obtained when we consider Neumann boundary condition, as it should be. In the next section, we will consider a mix of these two boundary conditions in which case we will refer to as mixed boundary condition. 
%
\subsection{Mixed boundary condition}
%
After the analysis of the effective potential, Casimir-like effect and topological mass assuming Dirichlet and Neumann boundaries conditions obeyed by a massive scalar field on two  parallel plates, now we want to consider the case in which the field satisfies mixed boundary condition. In other words, we assume that the field obeys Dirichlet and Neumann boundary conditions on each of the plates separately. The conditions are then written as
\begin{equation}
 \label{mixed_cond}
 \begin{aligned}
  \varphi(x)\big{|}_{z=0}=\frac{\partial \varphi(x)}{\partial z}\bigg{|}_{z=a}=0 \ \ \ \ \ \text{and} \ \ \ \ \ \frac{\partial \varphi(x)}{\partial z}\bigg{|}_{z=0}=\varphi(x)\big{|}_{z=a} \ .
 \end{aligned}
\end{equation}
The complete set of normalized solutions of the scalar field, $\varphi$, under these conditions have also been reported in Ref. \cite{cruz:2017}.  These solutions provide the following eigenvalues of the operator \eqref{operator}:
\begin{eqnarray}
\Lambda_{\beta} = k_t^2 + k^2 + \left[\left(n+\frac{1}{2}\right)\frac{\pi}{a}\right]^2  -\chi u^{\mu}u^{\nu}k_{\mu}k_{\nu} + m^2 + \frac{\lambda\Phi^2}{2},
\label{evmb}
\end{eqnarray}
where $k^2=k_x^2+k_y^2$ and  $n=0,1,2,3,...$\ . Hence, we will consider the set of eigenvalues \eqref{evmb} representing the mixed boundary condition case to calculate the renormalized effective potential, the vacuum energy and topological mass taking into consideration the two types of Lorentz symmetry violation, namely, the timelike and spacelike types.
%
\subsubsection{Timelike vector}
%
In the case that the $4$-vector $u^{\mu}$ is of the timelike type, i.e., $ u^{t}=(1,0,0,0)$, the eigenvalues \eqref{evmb} of the elliptic operator \eqref{operator} becomes 
\begin{equation}
 \Lambda_{\beta} = \left(1+\chi \right)k_t^2 + k^2 + \left[\left(n+\frac{1}{2}\right)\frac{\pi}{a}\right]^2 + m^2 + \frac{\lambda \Phi^2}{2} \ .
\end{equation}
Consequently, the zeta function \eqref{zet_funct_DN_t_0} is now written as
\begin{equation}
 \zeta(s) = \frac{1}{\sqrt{1+\chi}} \frac{V_3}{(2\pi)^3} \sum_{n=0}^{\infty} \int d^3k \left \{ \kappa_t^2 + k^2 + \left[\left(n+\frac{1}{2}\right)\frac{\pi}{a}\right]^2 + m^2 + \frac{\lambda \Phi^2}{2} \right \}^{-s} \ .
\end{equation}
Furthermore, by using the identity \eqref{identity} we are able to perform the integrals in $\kappa_t$, $k_x$ and $k_y$, providing that
\begin{equation}
\label{zet_funct_M_t_0}
 \zeta(s) = \frac{1}{\sqrt{1+\chi}} \frac{V_3}{(2\pi)^3}\frac{\pi^{3/2}\Gamma(s-3/2)}{\Gamma(s)}w^{3-2s}\sum_{n=0}^{\infty}\left[\left(n+\frac{1}{2}\right)^2+\nu^2\right]^{3/2-s} \ ,
\end{equation}
where
\begin{equation}
 \begin{aligned}
  \nu^2 = \frac{\lambda \Phi^2}{2w^2} + \frac{m^2}{w^2} \ \ \ \ \ \ \text{and} \ \ \ \ \ \ \ w = \frac{\pi}{a} \ .
 \end{aligned}
\end{equation}
We can see that there is still a sum in $n$ to be performed in the zeta function expression \eqref{zet_funct_M_t_0}. In this sense, in order to apply the Epstein-Hurwitz zeta function  \eqref{EH_funct} we can write the sum in $n$ as
\begin{equation}
  \label{sum_alt_M_t}
  \sum_{n=0}^{\infty}\left[\left(n+\frac{1}{2}\right)^2+\nu^2\right]^{3/2-s} = \frac{1}{2^{3-2s}} \left[\sum_{n=1}^{\infty}\left(n^2+4\nu^2\right)^{3/2-s} - 2^{3-2s}\sum_{n=1}^{\infty}\left(n^2+\nu^2\right)^{3/2-s}\right] .
\end{equation}
By using now Eqs. \eqref{sum_alt_M_t} and \eqref{EH_funct} in \eqref{zet_funct_M_t_0} we have the final form of the generalized zeta function as given by
\begin{equation}
\label{zet_funct_M_t_f}
 \begin{aligned}
  \zeta(s) &= \frac{V_3}{(2\pi)^3} \Bigg{\{}\frac{\pi^2w^{-2\bar{s}}\nu^{1-2\bar{s}}\Gamma(\bar{s}-1/2)}{2\Gamma(\bar{s}+3/2) \sqrt{1+\chi}} + \frac{2^{\bar{s}+1/2}\pi^{2\bar{s}+1}w^{-2\bar{s}}}{\Gamma(\bar{s}+3/2) \sqrt{1+\chi}} \times \\ & \ \ \ \ \ \ \ \ \ \ \ \ \ \times \Bigg{[}4^{\bar{s}} \sum_{n=1}^{\infty} n^{2\bar{s}-1}f_{\bar{s}-1/2}\big(4\pi n \nu \big) - \sum_{n=1}^{\infty} n^{2\bar{s}-1}f_{\bar{s}-1/2}\big(2\pi n \nu \big) \Bigg{]} \Bigg{\}} \ ,
 \end{aligned}
\end{equation}
where $\bar{s}=s-3/2$. Thus, by using \eqref{zet_funct_M_t_f} we can obtain $\zeta(0)$ and $\zeta^{\prime}(0)$ leading to the one-loop correction to the effective potential \eqref{one_loop_potential_DN_t} written as
\begin{equation}
\label{one_loop_poten_M_t}
 \begin{aligned}
  V^{(1)}(\Phi) &= - \frac{3 b^4}{128 \pi^2 \sqrt{1+\chi}} + \frac{b^4 \text{ln}\left(\frac{b^2}{\mu^2}\right)}{64 \pi^2 \sqrt{1+\chi}}- \frac{b^2}{16 \pi^2 \sqrt{1+\chi} a^2} \times \\ & \ \ \ \times \sum_{n=1}^{\infty} \frac{\left[K_2\left(4ban\right) - 2K_2\left(2ban\right)\right]}{n^2} \ ,
 \end{aligned}
\end{equation}
where
\begin{equation}
 b = \sqrt{\frac{\lambda \Phi^2}{2} + m^2} \ .
\end{equation}
Hence, from Eqs. \eqref{potencial_u} and \eqref{one_loop_poten_M_t} the effective potential up to one-loop correction is found to be
\begin{equation}
\label{effec_potent_M_t}
 \begin{aligned}
  V_{\text{eff}}(\Phi) &= \frac{m^2\Phi^2}{2} + \frac{\lambda \Phi^4}{4!} + \frac{\Phi^2}{2}\delta_2 + \frac{\Phi^4}{4!}\delta_1 + \delta_3 - \frac{3 b^4}{128 \pi^2 \sqrt{1+\chi}} + \frac{b^4 \text{ln}\left(\frac{b^2}{\mu^2}\right)}{64 \pi^2 \sqrt{1+\chi}} \\ & \ \ \  - \frac{b^2}{16 \pi^2 \sqrt{1+\chi} a^2} \sum_{n=1}^{\infty} \frac{\left[K_2\left(4ban\right) - 2K_2\left(2ban\right)\right]}{n^2} \ ,
 \end{aligned}
\end{equation}
where the renormalization constants $\delta_1$, $\delta_2$ and $\delta_3$ need to be determined in order to find the renormalized form for the effective potential \eqref{effec_potent_M_t}. For this purpose, the conditions \eqref{ren1}, \eqref{ren2} and \eqref{ren3} provide
\begin{equation}
 \frac{\delta_1}{4!} = \frac{\lambda^2 \text{ln} \left(\frac{\mu^2}{m^2}\right)}{256 \pi^2 \sqrt{1+\chi}}\ ,
\end{equation}
\begin{equation}
  \frac{\delta_2}{2} = \frac{\lambda m^2 \text{ln} \left(\frac{\mu^2}{m^2}\right)}{64 \pi^2
   \sqrt{1+\chi}} + \frac{\lambda 
   m^2}{64 \pi^2 \sqrt{1+\chi}}\ ,
\end{equation}
and
\begin{equation}
 \delta_3 = \frac{m^4 \text{ln} \left(\frac{\mu^2}{m^2}\right)}{64 \pi^2 \sqrt{1+\chi}} + \frac{3 m^4}{128 \pi^2 \sqrt{1+\chi}}\ .
\end{equation}
The renormalization constants found above, when used in \eqref{effec_potent_M_t}, allow us to obtain the renormalized effective potential up to one-loop correction
\begin{equation}
\label{ren_effec_potent_M_t}
 \begin{aligned}
  V_{\text{eff}}^{\text{R}}(\Phi) &= \frac{m^2 \Phi^2}{2} + \frac{\lambda  \Phi^4}{24} -\frac{3 \lambda^2 \Phi^4}{512 \pi ^2 \sqrt{1+\chi}} - \frac{\lambda  m^2 \Phi^2}{128 \pi^2 \sqrt{1+\chi}} \\ & \ \ + \frac{\lambda^2 \Phi^4 \text{ln} \left(\frac{b^2}{m^2}\right)}{256 \pi^2 \sqrt{1+\chi}}  + \frac{\lambda m^2 \Phi^2 \text{ln} \left(\frac{b^2}{m^2}\right)}{64 \pi^2 \sqrt{1+\chi}} + \frac{m^4 \text{ln} \left(\frac{b^2}{m^2}\right)}{64 \pi^2 \sqrt{1+\chi}} \\ & \ \  - \frac{b^2}{16 \pi^2 \sqrt{1+\chi } a^2} \sum_{n=1}^{\infty} \frac{\left[K_2\left(4abn\right)-2K_2\left(2abn\right)\right]}{n^2} \ .
 \end{aligned}
\end{equation}
The renormalized effective potential above, at $\Phi=0$, provide the vacuum energy per unit area of the plates as
\begin{equation}
\label{cas_ener_dens_M_t_ex}
 \begin{aligned}
  \frac{E_C}{L^2} = a V_{\text{eff}}^{\text{R}}(0) = - \frac{m^2}{16 \pi^2 \sqrt{1+\chi } a} \sum_{n=1}^{\infty} \frac{\left[K_2\left(4amn\right)-2K_2\left(2amn\right)\right]}{n^2} \ .
 \end{aligned}
\end{equation}
This expression is a convergent and exact expression for the vacuum energy. From it we can consider asymptotic expressions for small and large arguments of the modified Bessel function $K_{\mu}(x)$. 

Let us now show the asymptotic expressions in the regimes $ma\ll 1$ and $ma\gg 1$. In the latter, the vacuum energy \eqref{cas_ener_dens_M_t_ex} is exponentially suppressed and dominated by the term $n=1$ of the modified Bessel function in the sum, i.e.,
\begin{equation}
  \frac{E_C}{L^2} \approx \frac{1}{32 \sqrt{1+\chi}}\left(\frac{m}{\pi a}\right)^{3/2} e^{-2am} \ .
 \end{equation}
On the other hand, in the regime $ma\ll 1$, the vacuum energy is given by
 \begin{equation}
  \begin{aligned}
   \frac{E_C}{L^2} \approx \frac{7 \pi^2}{11520 \sqrt{1+\chi}a^3} -\frac{a m^4}{48 \pi^2 \sqrt{1+\chi}} + \frac{a^2 m^5}{60 \pi^2 \sqrt{1+\chi}} \ .
     \label{Eap}
  \end{aligned}
 \end{equation}
This approximated expression is dominated by the first term on the r.h.s, associated with the massless scalar field. 

Now we turn to the calculation of the two-loop correction to the effective potential. As in the previous sections, we can also make use of the zeta function which in the present case is given by \eqref{zet_funct_M_t_f}. Thus, the function $S_1(\Phi)$ is written in the form
\begin{equation}
 S_1(\Phi) = \left \{ \sum_{n=0}^{\infty} \frac{1}{a} \int \frac{d^3k}{(2\pi)^3}\left[\left(1+\chi \right)k_t^2 + k^2 + \left(\left(n+\frac{1}{2}\right)\frac{\pi}{a}\right)^2 + m^2 + \frac{\lambda \Phi^2}{2}\right]^{-s}\right \}^2 \ ,
\end{equation}
and can be expressed in terms of the zeta function \eqref{zet_funct_M_t_f} as
\begin{equation}
 S_1(\Phi) =\left[\frac{\zeta_R(1)}{V_3a}\right]^2,
 \label{s11}
\end{equation}
where $\zeta_R(1)$ is the zeta function \eqref{zet_funct_M_t_f} taken at $s = 1$ after subtracting the divergent part of it given by the first term on the r.h.s. As explained before, this divergent part, when divided by $V_3a$, does not depend on $a$ and as customary must be subtracted. Thus, from \eqref{s11} and \eqref{vac_two_loop_contr_potent}, we obtain the two-loop correction to the vacuum energy as
\begin{equation}
 \label{2_loop_cas_energ_M_t}
\frac{E^{(\lambda)}_C}{L^2} = a V^{(2)}(0) = \frac{\lambda m^2}{128 \pi^4 (1+\chi) a} \left \{ \sum_{n=1}^{\infty}\frac{\left[K_1(4amn) - K_1(2amn)\right]}{n} \right \}^2 \ ,
\end{equation}
which is also a convergent and exact expression. It is also exponentially suppressed for $ma\gg 1$ and provide the massless contribution for $ma\ll 1$. 

The exponentially suppressed mathematical expression for \eqref{2_loop_cas_energ_M_t} in the regime $ma\gg 1$ is given by
\begin{equation}
\frac{E^{(\lambda)}_C}{L^2} \approx \frac{\lambda m e^{-4 a m}}{512 \pi^3 \left(1+\chi \right) a^2} \ ,
 \end{equation}
while in the regime $ma\ll 1$, the vacuum energy is written as
\begin{equation}
\frac{E^{(\lambda)}_C}{L^2} \approx \frac{\lambda}{73728 (1+\chi) a^3} - \frac{\lambda m^2}{3072 \pi^2 (1+\chi) a} + \frac{\lambda m^3}{4608 \pi^2 (1+\chi)} \ ,
 \label{asysmall}
\end{equation}
where we can clearly see that the first term on the r.h.s is the dominant one and is associated with the massless scalar field. 

\begin{figure}[h!]
    \centering
    \subfloat{{\includegraphics[width=17cm]{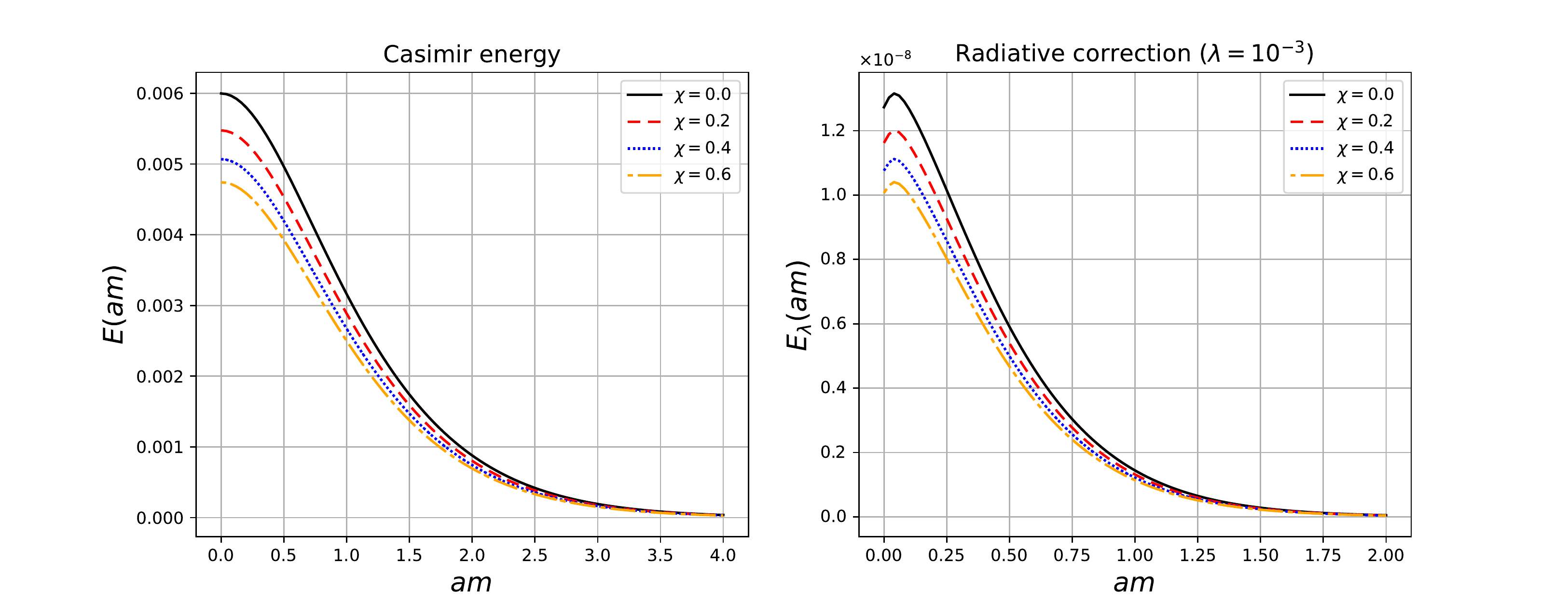} }}
    \qquad
    \caption{The behavior of the Casimir energy per unity area $E(am)=\frac{a^3}{L^2}E_C$ given by \eqref{cas_ener_dens_M_t_ex} as function of $am$ is exhibited in the left plot. The behaviour of the two-loop contribution $E_{\lambda}(am)=\frac{a^3}{L^2}E^{(\lambda)}_C$ , given by \eqref{2_loop_cas_energ_M_t}, is exhibited in the right plot. For the latter we assume $\lambda=10^{-3}$, and consider different values for the parameter, $\chi$.}
    \label{figure6}
\end{figure}

In the left panel of Fig.\ref{figure6} we exhibit the Casimir energy, given by \eqref{cas_ener_dens_M_t_ex}, as function of $am$, wheres in the right panel we exhibit the behavior of  the two-loop correction to the Casimir energy per unity area, given by \eqref{2_loop_cas_energ_M_t},  as function of $am$ considering different values for $\chi$ and fixing $\lambda=10^{-5}$. By these plots we can infer that the vacuum energy and its two-loop correction decrease as $\chi$ increases. Note that this is different from the timelike case considering only Dirichlet/Neumann boundary condition, in which case the vacuum energy increases whereas its radiative correction decreases, as $\chi$ increases.

The mixed boundary condition we are considering here will also generate, at one-loop level, a topological mass. In this sense, we can obtain the topological mass by using \eqref{ren_effec_potent_M_t} and \eqref{ren2}. It is given by 
\begin{equation}
\label{topol_mass_M_t}
  m_{\text{T}}^2 = m^2 \left \{ 1 + \frac{\lambda}{8 \pi^2 \sqrt{1+\chi} a m} \sum_{n=1}^{\infty} \frac{\left[K_1(4amn) - K_1(2amn) \right]}{n} \right \} \ .
\end{equation}

\begin{figure}[h!]
	\centering
	\includegraphics[scale=0.5]{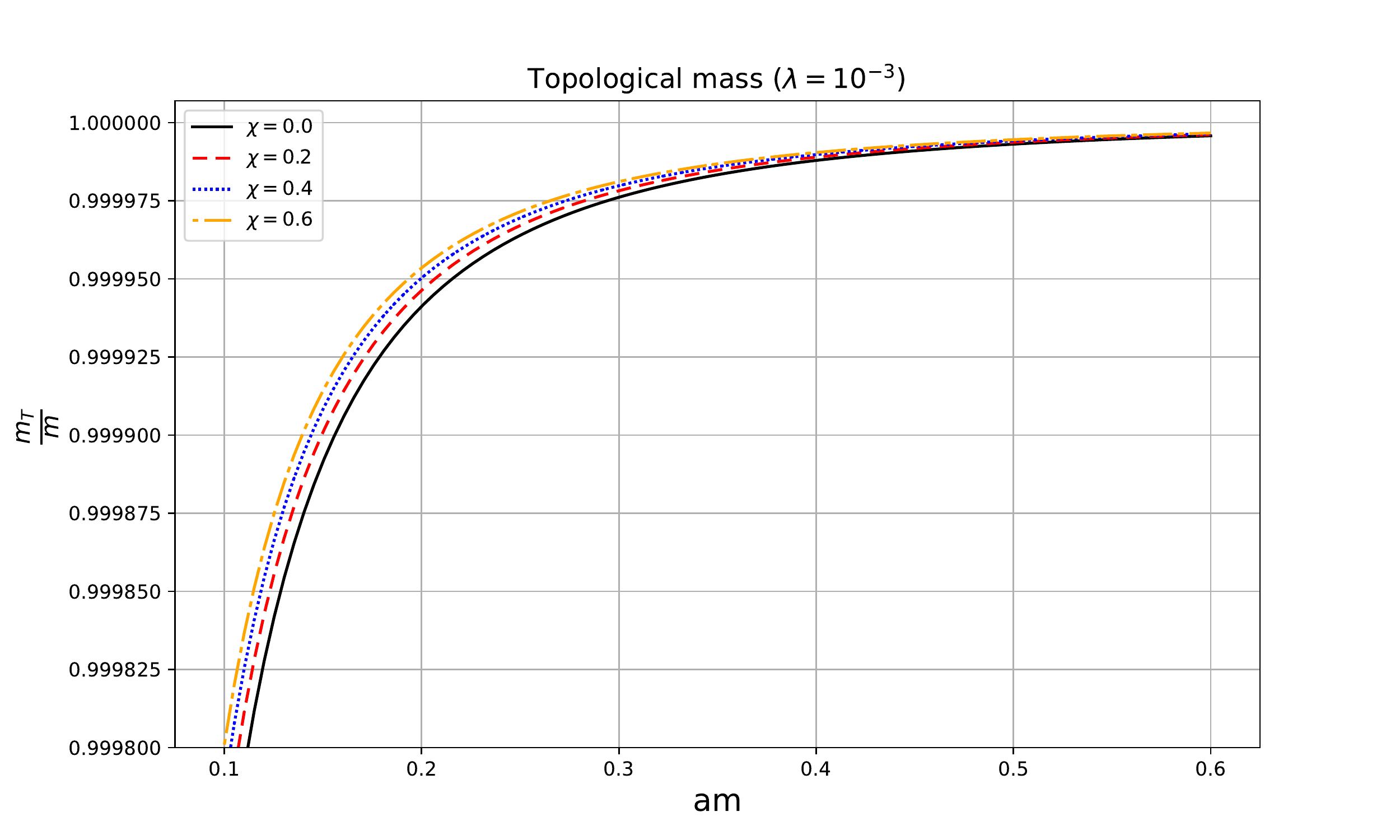}
	\caption{The ration of the topological mass by the field mass as function of $am$. In the plot is consider $\lambda=10^{-3}$ and different values of $\chi$.}
	\label{figure7}
\end{figure}

We have plotted in Fig.\ref{figure7} the behaviour of $\frac{m_T}{m}$ by using \eqref{topol_mass_M_t} in terms of $am$. The plot shows that the topological mass increases as the Lorentz symmetry violation parameter, $\chi$, increases. This is different from the timelike case considering only Dirichlet/Neumann boundary condition., in which case the topological mass decreases as $\chi$ increases. Fig.\ref{figure7} also shows that in the regime $ma\gg 1$ the topological mass is dominated by the first term on the r.h.s of \eqref{topol_mass_M_t} while in the opposite limit $ma\ll 1$ the topological mass is the one associated with a massless scalar field. 

Mathematically, for $ma\gg 1$, we have
 \begin{equation}
  m_{\text{T}}^2 \approx m^2 - \frac{\lambda}{16\sqrt{1+\chi}}\frac{\sqrt{ m} e^{-2 a m}}{\left(\pi a\right)^{3/2} } \ ,
 \end{equation}
which shows that the topological mass is dominated by the first term, $m^2$. 

In the opposite regime, $ma\ll 1$ we obtain
\begin{equation}
  m_{\text{T}}^2 \approx m^2 - \frac{\lambda}{192 \sqrt{1+\chi} a^2} + \frac{\lambda m^2}{16 \pi^2 \sqrt{1+\chi}} - \frac{\lambda a m^3}{24 \pi^2 \sqrt{1+\chi}} + \frac{\lambda a^3 m^5}{120 \pi^2 \sqrt{1+\chi}} \ .
 \end{equation}
In this limit, the dominant term is the second one on the r.h.s of the above approximation, associated with a massless scalar field.

\subsubsection{Spacelike vector}
%
We want now to consider the case in which the constant 4-vector, $u^{\mu}$, is of the spacelike type. As before, there are three spacelike components specifying the broken symmetry direction: $u^x=(0,1,0,0)$, $u^y=(0,0,1,0)$,  both parallel to the plates, and $u^z=(0,0,0,1)$, orthogonal to the plates. In the two first cases, specifying the $x$ and $y$ directions, parallel to the plates, the results for the effective potential, Casimir energy and topological mass are the same. Let us then consider the $x$-direction:
\begin{equation}
 u^x = (0,1,0,0) \ .
\end{equation}
In this case, from Eq. \eqref{evmb}, the set of eigenvalues is given by
\begin{equation}
 \label{eigev_M_x}
 \Lambda_{\beta} = k^2 + \left(1-\chi \right)k_x^2 + \left[\left(n+\frac{1}{2}\right)\frac{\pi}{a}\right]^2 + m^2 + \frac{\lambda \Phi^2}{2} \ ,
\end{equation}
where $k^2=k_t^2+k_y^2$. So, using Eq. \eqref{zet_funct_DN_t_0}, the zeta function is written as
\begin{equation}
 \zeta(s) = \frac{V_3}{(2\pi)^3} \int d^3k \sum_{n=1}^{\infty}\left \{ k^2 + \left(1-\chi \right)k_x^2 + \left[\left(n+\frac{1}{2}\right)\frac{\pi}{a}\right]^2 + m^2 + \frac{\lambda \Phi^2}{2}\right \}^{-s} \ .
\end{equation}
Again, by using the identity \eqref{identity}, we are able to solve the integrals in $k_t$, $k_x$ and $k_y$ to obtain the zeta function in the form
\begin{equation}
\label{zet_funct_M_x_0}
 \zeta(s) = \frac{1}{\sqrt{1-\chi}} \frac{V_3}{(2\pi)^3}\frac{\pi^{3/2}\Gamma(s-3/2)}{\Gamma(s)}w^{3-2s}\sum_{n=0}^{\infty}\bigg[\bigg(n+\frac{1}{2}\bigg)^2+\nu^2\bigg]^{3/2-s} \ ,
\end{equation}
where
\begin{equation}
 \nu^2 = \frac{\lambda \Phi^2}{2w^2}+\frac{m^2}{w^2} \ \ \ \ \ \ \ \text{and} \ \ \ \ \ \ \ w=\frac{\pi}{a} \ .
\end{equation}
We can notice that the expression \eqref{zet_funct_M_x_0} can be obtained from \eqref{zet_funct_M_t_0} by making $\chi \rightarrow -\chi$. Consequently, all the results for the renormalized effective potential, Casimir energy and topological mass can be obtained from the timelike case considered previously. This is also valid if we consider the vector $u^{y}$.

The most important type of Lorentz symmetry violation in this section is the one occurring in the orthogonal direction to the plates, that is,
\begin{equation}
 u^z = (0,0,0,1) \ .
\end{equation}
In this case, the set of eigenvalues \eqref{evmb} becomes
\begin{equation}
\label{eigev_M_z}
 \Lambda_{\beta} = k^2 + \left(1-\chi \right)  \left[\left(n+\frac{1}{2}\right)\frac{\pi}{a}\right]^2 + m^2 + \frac{\lambda \Phi^2}{2} \ ,
\end{equation}
where  $k^2=k_t^2+k_x^2+k_y^2$. Thus, substituting Eq. \eqref{eigev_M_z} in Eq. \eqref{zet_funct_DN_t_0}, we have the zeta function expression written as
\begin{equation}
 \label{zet_0_M_z}
 \zeta(s) = \frac{V_3}{(2\pi)^3} \sum_{n=0}^{\infty} \int d^3k \left \{ k^2 + \left(1-\chi \right)\left[\left(n+\frac{1}{2}\right)\frac{\pi}{a}\right]^2 + m^2 + \frac{\lambda \Phi^2}{2}\right \}^{-s} \ .
\end{equation}
Once again, by using \eqref{identity}, we obtain
\begin{equation} \label{zeta_func_mixed_z}
 \begin{aligned}
  \zeta(s)=\frac{V_3}{(2\pi)^3}\frac{\pi^{3/2}\Gamma(s-3/2)}{\Gamma(s)}w^{3-2s}\sum_{n=0}^{\infty}\bigg[\bigg(n+\frac{1}{2}\bigg)^2+\nu^2\bigg]^{3/2-s} \ ,
 \end{aligned}
\end{equation}
where
\begin{equation}
 \begin{aligned}
  \nu^2 \equiv \frac{\lambda \Phi^2}{2w^2} + \frac{m^2}{w^2} \ \ \ \ \ \ \text{and} \ \ \ \ \ \ \ w \equiv \sqrt{1-\chi}\frac{\pi}{a} \ .
 \end{aligned}
\end{equation}
We need now to find an expression for the sum in $n$ present in the zeta function above. In order to do that, we can make use of Eq. \eqref{sum_alt_M_t}. This provides
\begin{equation}
 \label{zeta_func_mixed_z_f}
  \zeta(s) = \frac{V_3}{(2\pi)^3}\frac{\pi^{3/2} \Gamma(\bar{s})}{\Gamma(\bar{s}+3/2)} \left(\frac{w}{2}\right)^{-2\bar{s}} \left[ \sum_{n=1}^{\infty}\left(n^2+4\nu^2\right)^{-\bar{s}} - 2^{-2\bar{s}}\sum_{n=1}^{\infty} \left(n^2+\nu^2\right)^{-\bar{s}}\right] \ ,
\end{equation}
where we have $\bar{s}=s-3/2$. Hence, the Epstein-Hurwitz zeta function \eqref{EH_funct} allows us to obtain \eqref{zeta_func_mixed_z_f} as
\begin{equation} \label{zeta_func_fin_mix_z}
 \begin{aligned}
  \zeta(s) &= \frac{V_3}{(2\pi)^3} \Bigg{\{}\frac{\pi^2w^{-2\bar{s}}\nu^{1-2\bar{s}}\Gamma(\bar{s}-1/2)}{2\Gamma(\bar{s}+3/2)} + \frac{2^{\bar{s}+1/2}\pi^{2\bar{s}+1}w^{-2\bar{s}}}{\Gamma(\bar{s}+3/2)} \\ & \ \ \ \ \ \ \ \ \ \ \ \ \times \left[4^{\bar{s}} \sum_{n=1}^{\infty} n^{2\bar{s}-1}f_{\bar{s}-1/2}\left(4\pi \nu n \right) - \sum_{n=1}^{\infty} n^{2\bar{s}-1}f_{\bar{s}-1/2}\left(2\pi \nu n \right) \right] \Bigg{\}} \ .
 \end{aligned}
\end{equation}
Consequently, in the limit $s\rightarrow 0$, we have
\begin{equation} \label{zeta_mix_0_z}
  \zeta(0) = \frac{V_3 a}{2(2\pi)^4} \frac{\pi^2 b^4}{\sqrt{1-\chi}} \ ,
\end{equation}
and
\begin{equation} \label{zeta_prim_mix_0_z}
 \begin{aligned}
  \zeta^{\prime}(0) = \frac{V_3 a}{(2\pi)^3} \Bigg{\{} & \frac{3 \pi b^4}{8 \sqrt{1-\chi}} - \frac{\pi b^4 \text{ln}(b)}{2 \sqrt{1-\chi}} \\ & + \frac{\pi b^2 \sqrt{1-\chi}}{a^2} \sum_{n=1}^{\infty}\frac{\left[ K_2 \left(\frac{4abn}{\sqrt{1-\chi}}\right) - 2K_2\left(\frac{2abn}{\sqrt{1-\chi}}\right)\right]}{n^2} \Bigg{\}} \ .
  \end{aligned}
\end{equation}
The one-loop correction \eqref{one_loop_potential_DN_t} to the effective potential is now possible to be obtained by using the results above for $\zeta(0)$ and $\zeta'(0)$. This gives
\begin{equation}
 \label{1_loop_potent_M_z}
 \begin{aligned}
  V^{(1)}(\Phi) = & \frac{b^4 \text{ln}\left(\frac{b^2}{\mu^2}\right)}{64 \pi^2 \sqrt{1-\chi}} - \frac{3 b^4}{128 \pi^2 \sqrt{1-\chi}}  \\ & - \frac{b^2 \sqrt{1-\chi}}{16 \pi^2 a^2} \sum_{n=1}^{\infty} \frac{\left[K_2\left(\frac{4abn}{\sqrt{1-\chi}}\right) - 2K_2\left(\frac{2abn}{\sqrt{1-\chi}}\right)\right]}{n^2} \ ,
 \end{aligned}
\end{equation}
where
\begin{equation}
 b = \sqrt{\frac{\lambda \Phi^2}{2} + m^2} \ .
\end{equation}
Furthermore, the effective potential up to one-loop correction, from Eqs. \eqref{potencial_u} and Eq. \eqref{1_loop_potent_M_z}, is written as
\begin{equation}
\label{effec_potent_M_z}
 \begin{aligned}
  V_{\text{eff}}(\Phi) = & \frac{m^2\Phi^2}{2} + \frac{\lambda \Phi^4}{4!} + \frac{ \Phi^2}{2}\delta_2 + \frac{\Phi^4}{4!}\delta_1 + \delta_3+ \frac{b^4 \text{ln}\left(\frac{b^2}{\mu^2}\right)}{64 \pi^2\sqrt{1-\chi }} - \frac{3 b^4}{128 \pi^2\sqrt{1-\chi }} \\ &  - \frac{b^2 \sqrt{1-\chi }}{16 \pi^2 a^2} \sum_{n=1}^{\infty} \frac{\left[K_2\left(\frac{4abn}{\sqrt{1-\chi}}\right)  - 2K_2\left(\frac{2abn}{\sqrt{1-\chi}}\right)\right]}{n^2} \ ,
 \end{aligned}
\end{equation}
where the renormalization constants $\delta_1$, $\delta_2$ and $\delta_3$ are to be found by using \eqref{effec_potent_M_z} and the conditions given by $\eqref{ren1}$, $\eqref{ren2}$ and $\eqref{ren3}$. This provides 
\begin{equation}
\label{lambda_const_M_t}
  \frac{\delta_1}{4!} = \frac{\lambda^2 \text{ln} \left(\frac{\mu^2}{m^2}\right)}{256 \pi^2
   \sqrt{1-\chi}} 
\end{equation}
\begin{equation}
\label{m2_const_M_t}
  \frac{\delta_2}{2} =\frac{\lambda m^2 \text{ln} \left(\frac{\mu^2}{m^2}\right)}{64 \pi^2 \sqrt{1-\chi}} +\frac{\lambda m^2}{64 \pi^2 \sqrt{1-\chi}} \ ,
\end{equation}
and
\begin{equation}
\label{m4_const_M_t}
  \delta_3 = \frac{m^4 \text{ln} \left(\frac{\mu^2}{m^2}\right)}{64 \pi^2
   \sqrt{1-\chi}} + \frac{3
   m^4}{128 \pi^2 \sqrt{1-\chi}}\ .
\end{equation}
Finally, by using the renormalization constants found above in Eq. \eqref{effec_potent_M_z} we obtain the renormalized effective potential up to one-loop correction, i.e.,
\begin{equation}
\label{ren_effec_potent_M_z}
 \begin{aligned}
  V^{\text{R}}_{\text{eff}}(\Phi) = & \frac{m^2 \Phi^2}{2} + \frac{\lambda \Phi^4}{24} -\frac{\lambda m^2 \Phi^2}{128 \pi^2 \sqrt{1-\chi}} - \frac{3 \lambda^2 \Phi^4}{512 \pi^2 \sqrt{1-\chi}}\\ &+\frac{\lambda m^2 \Phi^2 \text{ln} \left(\frac{b^2}{m^2}\right)}{64 \pi^2 \sqrt{1-\chi}} + \frac{m^4 \text{ln} \left(\frac{b^2}{m^2}\right)}{64 \pi^2 \sqrt{1-\chi}} + \frac{\lambda^2 \Phi^4 \text{ln} \left(\frac{b^2}{m^2}\right)}{256 \pi^2 \sqrt{1-\chi}}   \\ & - \frac{b^2 \sqrt{1-\chi}}{16 \pi^2 a^2} \sum_{n=1}^{\infty} \frac{\left[K_2\left(\frac{4abn}{\sqrt{1-\chi}}\right)  - 2K_2\left(\frac{2abn}{\sqrt{1-\chi}}\right)\right]}{n^2} \ .
 \end{aligned}
\end{equation}

At this point we can, by using the renormalized effective potential \eqref{ren_effec_potent_M_z}, obtain the vacuum energy per unit area of the plate. This is done taking \eqref{ren_effec_potent_M_z} at $\Phi=0$. This gives
\begin{equation}
\label{casim_energ_dens_M_z}
  \frac{E_C}{L^2} = aV_{\text{eff}}^{\text{R}}(0) = - \frac{m^2 \sqrt{1-\chi }}{16 \pi^2 a} \sum_{n=1}^{\infty} \frac{\left[K_2\left(\frac{4amn}{\sqrt{1-\chi}}\right)  - 2K_2\left(\frac{2amn}{\sqrt{1-\chi}}\right)\right]}{n^2} \ .
\end{equation}
This exact and closed expression for the vacuum energy is exponentially suppressed for $ma\gg 1$ while for $ma\ll 1$ provides the expression for the vacuum energy in the massless scalar field case.

The exponentially suppressed expression for the vacuum energy in the regime $ma\gg 1$ is dominated by the $n=1$ term of the sum, providing
 \begin{equation}
   \frac{E_C}{L^2} \approx \frac{\left(1-\chi \right)^{3/4}}{16} \left( \frac{m}{\pi a}\right)^{3/2}  e^{-\frac{2 a m}{\sqrt{1-\chi}}} \ .
 \end{equation}
The opposite regime $ma\ll 1$ provides the approximated expression for the vacuum energy
 \begin{equation}
   \frac{E_C}{L^2} \approx  \frac{7 \pi^2 (1-\chi)^{3/2}}{11520 a^3} - \frac{a m^4}{48 \pi^2 \sqrt{1-\chi}} + \frac{a^2 m^5}{60 \pi^2 (1-\chi)} \ ,
   \label{masmall}
 \end{equation}
 which is dominated by the first term on the r.h.s. This is the term associated with the vacuum energy of the massless scalar field.  
\begin{figure}[h!]
	\centering
	\subfloat{{\includegraphics[width=17cm]{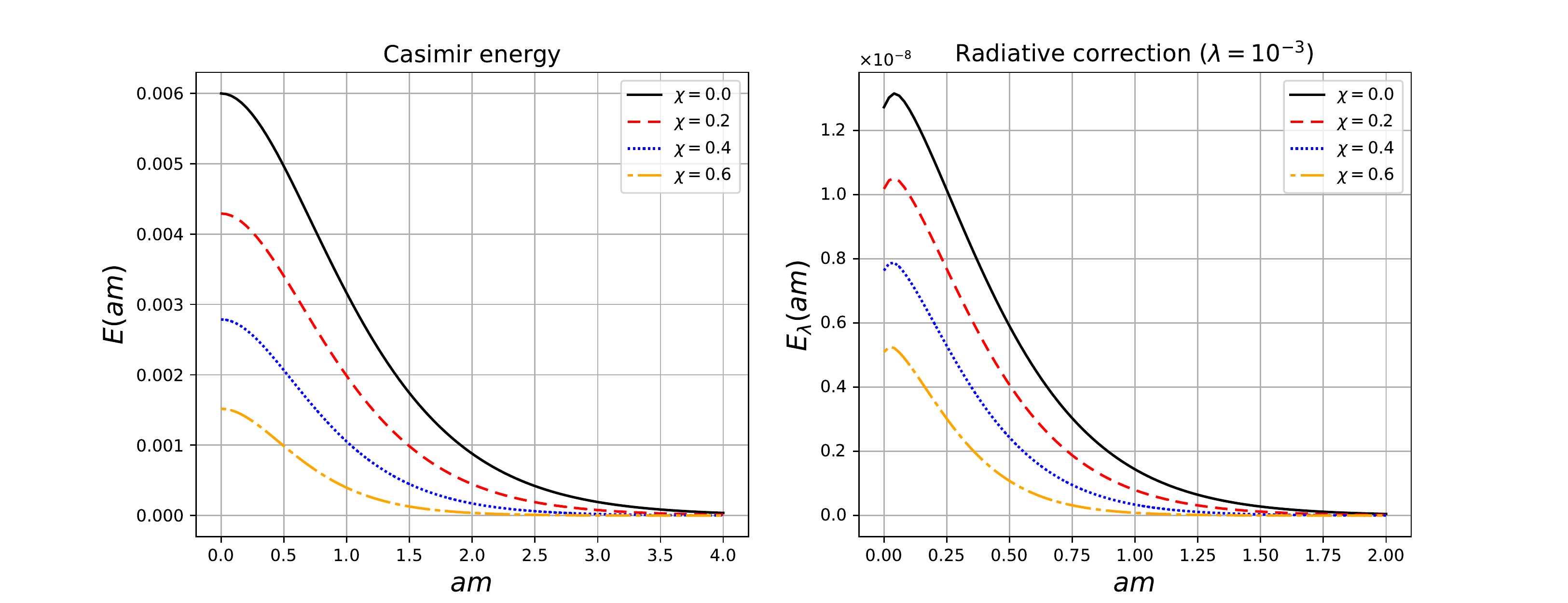} }}
	\qquad
	\caption{The left plot presents the behavior of Casimir energy per unity area $E(am)=\frac{a^3}{L^2}E_C$ , given by \eqref{casim_energ_dens_M_z}, as function of $am$. The two-loop contribution $E_{\lambda}(am)=\frac{a^3}{L^2}E^{(\lambda)}_C$, given by \eqref{2_loop_cas_energ_M_z}, is exhibited in the right panel as function of $am$ fixing $\lambda=10^{-3}$. In both plots we have considered $\chi$.}
	\label{figure8}
\end{figure}
 
 We can now turn the calculation of the two-loop correction to the effective potential at $\Phi=0$. The $S_1(\Phi)$ function, as before, is given by
\begin{equation} \label{S1_funct_M_z}
 S_1(\Phi) = \left \{ \sum_{n=0}^{\infty}\frac{1}{a} \int \frac{d^3k}{(2\pi)^3} \left[k^2 + \left(1-\chi \right)\left[\left(n+\frac{1}{2}\right)\frac{\pi}{a}\right]^2 + m^2 + \frac{\lambda \Phi^2}{2}\right]^{-s}\right \}^2 \ ,
\end{equation}
which can be expressed in terms of the zeta function \eqref{zeta_func_fin_mix_z} as
\begin{equation}
 S_1(\Phi) =\left[\frac{\zeta_R(1)}{V_3a}\right]^2,
 \label{s1mb}
\end{equation}
where $\zeta_R(1)$ is the zeta function \eqref{zet_funct_M_t_f} taken at $s = 1$ after subtracting the divergent part of it given by the first term on the r.h.s. Again, this divergent part, when divided by $V_3a$, does not depend on $a$ and, as customary, must be subtracted. Thus, from \eqref{s1mb} and \eqref{vac_two_loop_contr_potent}, we obtain the two-loop correction to the vacuum energy as
\begin{equation}
 \label{2_loop_cas_energ_M_z}
 \frac{E^{(\lambda)}_C}{L^2} = a V^{(2)}(0) = \frac{\lambda m^2}{128 \pi^4 a} \left \{ \sum_{n=1}^{\infty} \frac{\left[ K_1\left(\frac{4amn}{\sqrt{1-\chi}} \right) - K_1\left(\frac{2amn}{\sqrt{1-\chi}} \right)\right]}{n} \right \}^2 \ .
\end{equation}
This is an exact and convergent expression for the correction of the vacuum energy per unity area. The asymptotic behaviors for the above expression are explicitly provided below. 

The exponentially suppressed expression for the vacuum energy correction in the regime $ma\gg 1$ is dominated by the $n=1$ term: 
\begin{equation}
 \frac{E^{(\lambda)}_C}{L^2} \approx \frac{\lambda m  \sqrt{1-\chi} e^{-\frac{4 a
 m}{\sqrt{1-\chi}}}}{512 \pi^3 a^2} \ .
 \end{equation}
 On the other hand, the expression for the vacuum energy correction in the opposite regime $ma\ll 1$ is given by
  \begin{equation}
 \frac{E^{(\lambda)}_C}{L^2}  \approx \frac{\lambda (1-\chi)}{73728 a^3} - \frac{\lambda  m^2}{3072 \pi^2 a} +  \frac{\lambda m^3}{4608 \pi^2 \sqrt{1-\chi}} + \frac{\lambda a m^4}{512 \pi^4 (1-\chi)} \ .
 \label{ecmb}
 \end{equation}
 This expression is dominated by the first term on the r.h.s and is associated with the vacuum energy correction in the massless scalar field case. 
 
 In the left panel of Fig.\ref{figure8} we exhibit the behavior of the Casimir energy, \eqref{casim_energ_dens_M_z}, as function of $am$. In the right panel is exhibited the vacuum energy radiative correction, \eqref{2_loop_cas_energ_M_z}, also as function of $am$. In both cases, the energy values decreases  as $\chi$ increases. The curves are shifted down more than in the timelike case. This is due the dependence of the vacuum energy \eqref{casim_energ_dens_M_z} and its radiative correction \eqref{2_loop_cas_energ_M_z} on $\chi$, in the argument of the modified Bessel function, $K_{\mu}(x)$. The vacuum energy \eqref{casim_energ_dens_M_z}  also depend on, $\chi$, as a multiplicative factor.   
\begin{figure}[h!]
	\centering
	\includegraphics[scale=0.5]{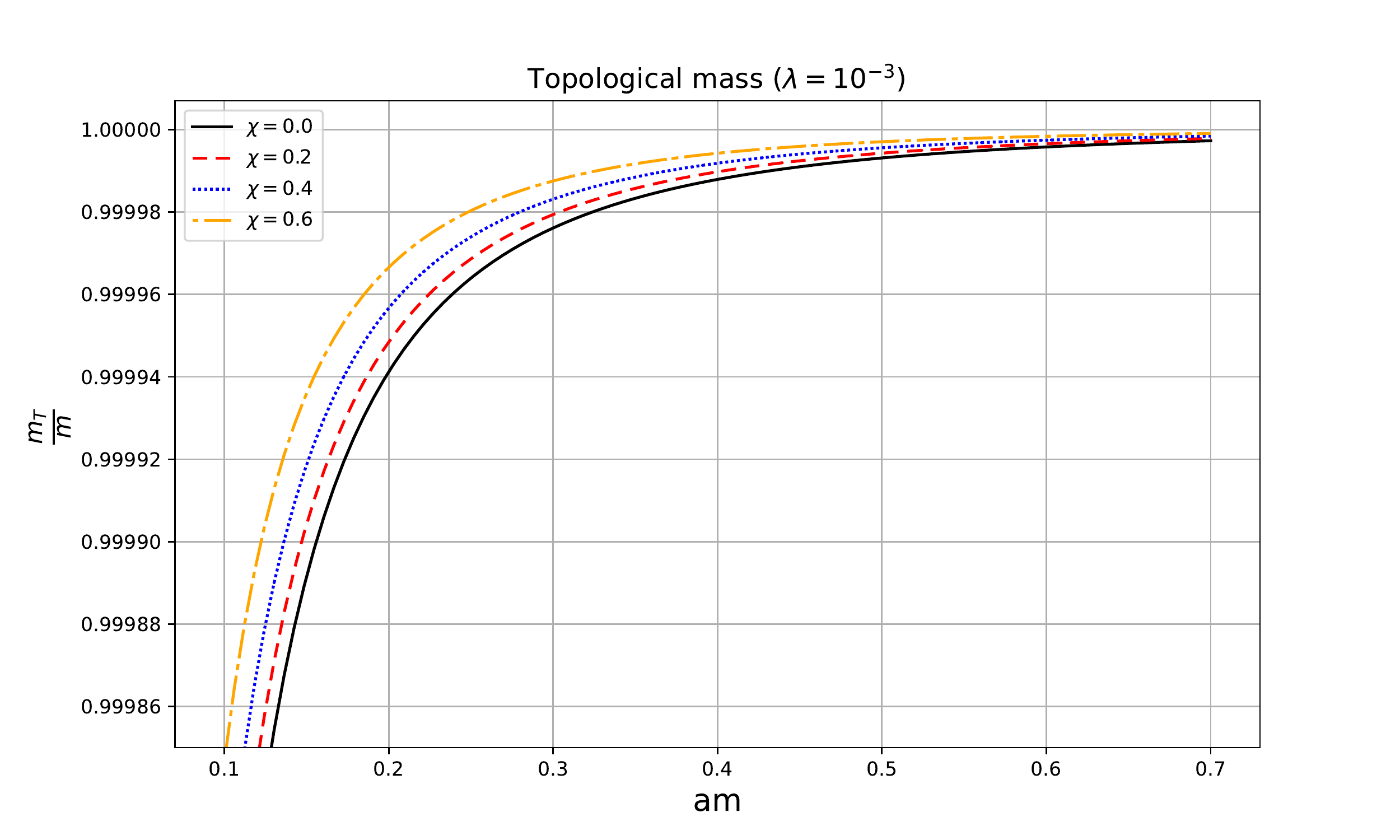}
	\caption{Graph presenting behavior of the ration of the topological mass by the mass of the field as function of $am$. In the plot is consider $\lambda=10^{-3}$ and different values for the Lorentz violating parameter: $\chi$.}
	\label{figure9}
\end{figure}

 We want now to analyse the generation of topological mass. The latter can be obtained by using the renormalized effective potential \eqref{ren_effec_potent_M_z} in the condition \eqref{ren2}. This provides the exact expression for the topological mass, that is,
\begin{equation}
 \label{topol_mass_M_z}
 m_{\text{T}}^2 = m^2 \left \{ 1 + \frac{\lambda}{8 \pi^2 a m} \sum_{n=1}^{\infty} \frac{\left[K_1\left(\frac{4amn}{\sqrt{1-\chi}} \right) - K_1\left(\frac{2amn}{\sqrt{1-\chi}} \right)\right]}{n} \right \} \ .
\end{equation}
This expression is plotted in Fig.\ref{figure9}. We can see that in the regime $ma\gg 1$ the topological mass is dominated by the first term on the r.h.s and grows to infinity. In the opposite regime $ma\ll 1$ the topological mass tends to the expression associated with the massless scalar field. We can also see in Fig.\ref{figure9} that the topological mass increases as $\chi$ increases. The curved are shifted up more than in the timelike previous case as a consequence of the dependence of the topological mass \eqref{topol_mass_M_z} on $\chi$, in the argument of the modified Bessel function.

The topological mass \eqref{topol_mass_M_z} in the regime $ma\gg 1$ is given by
\begin{equation}
  m_{\text{T}}^2 \approx m^2 - \frac{\lambda \sqrt[4]{1-\chi}}{16 \pi^{3/2}} \sqrt{\frac{m}{a^3}} e^{-\frac{2am}{\sqrt{1-\chi}}} \ ,
 \end{equation}
while in the opposite regime $ma\ll 1$ is
 \begin{equation}
  m_{\text{T}}^2 \approx m^2 - \frac{\lambda \sqrt{1-\chi}}{192 a^2} - \frac{\lambda  m^2}{16 \pi^2 \sqrt{1-\chi}} - \frac{\lambda a m^3}{24 \pi^2 (1-\chi)} + \frac{\lambda a^3 m^5}{120 \pi^2 (1-\chi)^2} \ .
  \label{tpsmall}
 \end{equation}
Note that the second term on the r.h.s of \eqref{tpsmall} is associated with a massless scalar field.

\section{Concluding remarks}
\label{concl}
.

In this work we have investigated the Casimir effect and the generation of topological mass associated with a scalar self-interacting, $\lambda \phi^4$, field theory in the context of aether-type Lorentz symmetry violation model, implemented by direct coupling between the derivative of the field with an external constant $4-$vector. Specifically we have considered the situation in which the field is confined between two parallel plates, assuming that it obeys, on each of the plates, Dirichlet, Newman and mixed boundary conditions, separately. The area of the plates has been taken to be $L^2$ wheres the distance between them has been taken as $a$ ($a \ll L$).

Furthermore, we have found exactly $\Phi$-dependent renormalized effective potentials, up to one loop-correction, considering both timelike and spacelike cases of the 4-vector $u^{\mu}$, where $\Phi$ is the classical and fixed background field. These renormalized effective potentials, at $\Phi=0$, provided a Casimir-like energy and topological mass for all cases. We have also obtained an exact two-loop correction to the effective potential when $\Phi=0$, which allows us to find a radiative correction to the Casimir-like energy obtained from the renormalized effective potential up to one-loop correction. The Casimir-like energies from Dirichlet and Neumann boundary conditions are equal and differ from the Casimir-like energy arising from the mixed boundary condition by a numerical factor and also by a change of sign. 

The Casimir-like effect, its radiative correction as well as the topological mass depend upon specific boundary conditions imposed on the fields and the Lorentz symmetry breaking parameter, $\chi$. It is worth pointing out that in all boundary condition cases considered, our results are more affected by $\chi$ in the spacelike type of broken symmetry, specifically, in the $z$-direction, orthogonal to the plates. Note that the results obtained here considering Dirichlet, Neumann and mixed boundary conditions, at one-loop level, agree with well known results in the case the Lorentz symmetry is preserved, that is, $\chi =0$ \cite{milton:2003, Bordag:2009zz}. This is also true at two-loop levels, i.e., we also recover, in the massless scalar field case, the expressions obtained for the Casimir energy density and topological mass in Ref. \cite{toms:1980:2805}, and the Casimir energy, considering the three boundaries conditions, in the massless field limit, given in  \cite{barone,barone1}. In all these three last papers, it was used the Riemann zeta-function renormalization to obtain the Casimir energy. The two-loop correction to the Casimir energy associated with the scalar field under Dirichlet boundary condition, in the absence of Lorentz symmetry violation, was also  calculated in \cite{reza}, using the Box Renormalization Scheme (BRS). However, in the latter, the Casimir energy correction in the massless limit disagrees with the results found in Refs.
\cite{toms:1980:2805,barone,barone1} by a negative sign. The reason for that is in the convention adopted in the definition of the two-loop correction in \cite{reza}, which presents a minus sign in the expression analogue to \ref{vac_two_loop_contr_potent} of our present paper. Consequently, the total Casimir energy will be decreased. From the physical point of view, it is expected that the scalar self-interaction would increase the Casimir energy, and not the opposite. Our results are in according with this assumption.

The analysis for the Casimir energy density considering an aether-type Lorentz symmetry violation term in the case of a Dirichlet self-interacting scalar field has been also considered in Ref. \cite{Mojavezi:2019ess}. There the  authors have obtained the first order radiative correction in $\lambda$ to the Casimir energy density by using BRS. Moreover, the definition adopted to evaluate this correction is the same as given in \cite{reza}, consequently a negative contribution to the Casimir energy is obtained. Finally we want to emphasize that in our present paper we have considered, besides Dirichlet  boundary condition, also Neumann and mixed ones to obtain the first order correction in $\lambda$ to the Casimir energy. Furthermore, we have used a different renormalization method based on the zeta function. We have also shown that in each one of the boundary conditions considered a topological mass is generated. 

Let us now, before ending the conclusions, discuss implications and the possibility of observational detection of a violation in the Lorentz symmetry in light of our results. As it is known the energy scale where the Lorentz symmetry is expected to be broken is of order of Planck scale, something around $10^{19}$\;GeV. This makes difficult in principle to envisage an experiment capable of detecting signals of Lorentz symmetry violation. Nevertheless, a Casimir energy density analysis considering models of Lorentz symmetry violation, as the one considered here, can offer a possible way of detecting signals in low energy scales. In particular, one can consider, for instance, extensions of the Standard Model of Particle Physics where violations in the Lorentz symmetry are taken into consideration. In these scenarios, looking at the Higgs sector where a beta decay is observed a bound of $\chi <10^{-6}$ is obtained. Also, a bound of $\chi <10^{-19}$ is obtained considering laser based on interferometry \cite{kostelecky:2003}. If these bounds are used in ours results finite values could be obtained and experiments for the detection of the Casimir energy could confirm the theoretical results. Likewise, if the detailed observations and measurements for the Casimir-like effect were possible, one could use the modifications of it by the Lorentz symmetry violation model considered here to estimate the values of the parameter, $\chi$, describing the spacetime anisotropy. This would certainly contribute to the experimental measurement attempts to get an upper bound on $\chi$.
\\
\\

{\bf Acknowledgements.} M.B.C is supported by Conselho
Nacional de Desenvolvimento Cient\'{\i}fico e Tecnol\'{o}gico - Brasil (CNPq) through the project No. 150479/2019-0. E.R.B.M is partially supported by CNPq under grant No 301.783/2019-3. H.F.S.M is partially supported  by CNPq under grants 305379/2017-8 and 430002/2018-1.

\end{document}